\documentclass[fleqn,usenatbib]{mnras}

\usepackage{newtxtext,newtxmath}

\usepackage[T1]{fontenc}
\usepackage{ae,aecompl}

\usepackage{changes}

\usepackage{graphicx}
\usepackage{epsfig}
\usepackage{epstopdf}
\usepackage{bmpsize}

\usepackage{graphicx}	

\title[Resolved star formation in galaxies at $z=1.5$ and $2.2$]
{A kpc-scale resolved study of unobscured and obscured star-formation activity in normal galaxies at {\it z} = 1.5 and 2.2 from ALMA and HiZELS
}

\author[C.~Cheng et al.]{Cheng~Cheng,$^{\! 1,2,3}$ 
Edo~Ibar,$^{\! 3}$
Ian~Smail,$^{\! 4}$
Juan~Molina,$^{\! 5}$
David~Sobral,$^{\! 6}$
Andr\'es~Escala,$^{\! 7}$
\newauthor
Philip~Best,$^{\! 8}$
Rachel~Cochrane,$^{\! 9,8}$
Steven~Gillman,$^{\! 4}$
Mark~Swinbank,$^{\! 4,10}$ 
R.\,J.~Ivison,$^{\! 11}$
\newauthor
Jia-Sheng~Huang,$^{\! 1,2}$
Thomas~M.~Hughes,$^{\! 1,3,12,13}$
Eric~Villard$^{\! 14}$
and Michele~Cirasuolo$^{\! 11}$
\\
$^{1}$Chinese Academy of Sciences South America Center for Astronomy, National Astronomical Observatories, CAS, Beijing 100101, China, \\
Email: chengcheng@nao.cas.cn\\
$^{2}$CAS Key Laboratory of Optical Astronomy, National Astronomical Observatories, Chinese Academy of Sciences, Beijing 100101, China, \\
$^{3}$Instituto de F\'isica y Astronom\'ia, Universidad de Valpara\'iso, Avda. Gran Breta\~na 1111, Valpara\'iso, Chile\\
$^{4}$Centre for Extragalactic Astronomy, Durham University, South Road, Durham, DH1~3LE\\
$^{5}$Kavli Institute for Astronomy and Astrophysics, Peking University, 5 Yiheyuan Road, Haidian District, Beijing 100871, P.R. China \\
$^{6}$Department of Physics, Lancaster University, Lancaster, LA1~4BY\\
$^{7}$Departamento de Astronom\'ia (DAS), Universidad de Chile, Casilla 36-D, Santiago, Chile\\
$^{8}$SUPA, Institute for Astronomy, Royal Observatory Edinburgh, EH9~3HJ\\
$^{9}$Harvard-Smithsonian Center for Astrophysics, 60 Garden St. Cambridge, MA 02138, USA\\
$^{10}$Institute for Computational Cosmology, Durham University, South Road, Durham DH1~3LE\\
$^{11}$European Southern Observatory, Karl-Schwarzschild-Strasse 2, D-85748 Garching, Germany\\
$^{12}$CAS Key Laboratory for Research in Galaxies and Cosmology, Department of Astronomy, University of Science and Technology of China, Hefei 230026, China\\
$^{13}$School of Astronomy and Space Science, University of Science and Technology of China, Hefei 230026, China\\
$^{14}$Joint ALMA Observatory/ESO Avenida Alonso de Cordova 3107, Vitacura, Santiago, Chile
}

\date{Accepted XXX. Received YYY; in original form ZZZ}

\pubyear{2020}

\begin{document}
\label{firstpage}
\pagerange{\pageref{firstpage}--\pageref{lastpage}}
\maketitle

\begin{abstract}
We present Atacama Large Millimeter/Submillimeter Array (ALMA) continuum observations
of a sample of nine star-forming galaxies at redshifts 1.47 and 2.23  selected
from the High-$z$ Emission Line Survey (HiZELS). Four galaxies in our sample are detected
at high significance by ALMA at a resolution of $0\farcs25$ at rest-frame 355\,$\mu$m.
Together with the previously observed H$\alpha$ emission, from adaptive optics-assisted 
integral-field-unit spectroscopy ($\sim0\farcs15$ resolution), and F606W and F140W imaging 
from the {\it Hubble Space Telescope} ($\sim0\farcs2$ resolution), we study the star-formation 
activity, stellar and dust mass in these high-redshift galaxies at $\sim$kpc-scale resolution. We
find that ALMA detection rates are higher for more massive galaxies
($M_*>10^{10.5}$\,M$_\odot$) and higher [N\,{\sc ii}]/H$\alpha$ ratios
($>0.25$, a proxy for gas-phase metallicity). 
The dust extends out to a radius of 8\,kpc, with a smooth structure, even for those galaxies 
presenting clumpy H$\alpha$ morphologies. The half-light radii ($R_{\rm dust}$) derived
for the detected galaxies are of the order $\sim$4.5\,kpc, more
than twice the size of submillimetre-selected galaxies at a similar
redshift. Our global star-formation rate estimates --- from far-IR and
extinction-corrected H$\alpha$ luminosities --- are in good agreement.
However, the different morphologies of the different phases of the interstellar 
medium suggest complex extinction properties
of the high-redshift normal galaxies.
\end{abstract}




\section{Introduction}

Observations of star-formation activity are critically important to
tackle open questions relating to galaxy formation and evolution. The
most direct view of cosmic star-formation history comes from observing
ultraviolet (UV) photons from the young massive stars. However, part
of this radiation is absorbed by dust, and the higher UV-energy
photons are absorbed by neutral hydrogen. Gas ionised by this
radiation eventually recombines, producing emission lines such as
Ly$\alpha$ and H$\alpha$, which have been used historically to
estimate the star-formation rate \citep[SFR, e.g., ][]{Sobral2019}.
Where the star-formation activity is shielded by gas and dust, the
ionising photons may get absorbed and re-radiated at far-infrared
(FIR) wavelengths by dust. 
Therefore, the rest-frame FIR emission is used to trace dust-obscured 
star formation in galaxies, and to derive total SFRs in combination 
with tracers of unobscured emission \citep[e.g.][]{Kennicutt1998, Kennicutt2012}.

Different ways to estimate the SFR in galaxies have their own
limitations and biases. For example, the H$\alpha$ recombination line
is generated originally by the photo-ionising radiation from massive
stars ($\gtrsim 10$\,M$_\odot$), or AGN, and is therefore sensitive to
recent star formation, within $\sim$10\,Myr, and has modest
sensitivity to dust obscuration. The UV flux ($\sim 1600$\,\AA), on
the other hand, comes from young, massive stars but is also emitted by
older 10--100\,Myr stars \citep{Kennicutt2012}, and is very sensitive
to dust obscuration. The FIR emission produced by heated dust grains
comes from young star-forming regions but in lower luminosity sources, 
it can also arise from dust heated by
older stellar populations.  Spatially-resolved observations of local
galaxies show that all of these SFR tracers are strongly correlated on
$\sim$\,kpc scales \citep[e.g.,][]{Boquien2016}.

Previous observations of high-redshift galaxies have shown that their
star-formation activity --- as traced by UV or H$\alpha$ emission ---
presents bright, clumpy star-forming complexes that could be up to
$\sim1000\times$ more massive than those seen in local galaxies
\citep[e.g., ][]{Genzel2006,Genzel2008, Schreiber2006, Schreiber2009,
  Schreiber2018, Shapiro2008, Swinbank2012a, Swinbank2012b}. However,
it may be that regions with fainter H$\alpha$ and UV emissions are 
affected by significant dust obscuration.  Previous work has shown
that dust attenuation causes the SFR derived from different indicators to be inconsistent
\citep{Swinbank2004, Katsianis2017}, thus it is
essential to have a spatially resolved view of the ionised gas and the
dust content in order to characterise the total SFR (unobscured and
obscured), especially for high-redshift galaxies.

\begin{table*}
\centering
\caption{\label{tab:table1} ALMA observational set-up, ordered by the observation date.}
\small
\begin{tabular}{lccccccccc}
\hline
Source List   & Project ID     & Observation    & Flux       & Bandpass   & Phase      & P.W.V.  & Number of & Band ($\nu_{\rm obs}$ /GHz) & Time on \\
              &                & Date           & Calibrator & Calibrator & Calibrator & (mm) & antennas    &    & Target (min)\\
\hline
SHiZELS-8     & 2012.1.00402.S &  6 Nov.\ 2013  & Uranus     & J2148+0657 & J0215-0222 & 3.9  & 29 &  7 (344)     & 52 \\
              &                &  29 Nov.\ 2013 & Uranus     & J0334-4008 & J0215-0222 & 1.3  & 26 &      & 43 \\
              &                &  29 Nov.\ 2013 & J0423-013  & J0334-4008 & J0215-0222 & 1.0  & 26 &       & 43 \\
\hline
SHIZELS-7     & 2013.1.01188.S & 11 Aug.\ 2015 & Ceres       & J0006-0623 & J0208-0047 & 0.9 & 43  &  7 (344)     & 28 \\
SHiZELS-21    &                & 29 Aug.\ 2015 & J0334-401   & J0006-0623 & J0219+0120 & 1.5 & 37  &  6 (261)     & 30 \\
SHiZELS-2     &                & 29 Aug.\ 2015 & J0238+166   & J0224+0659 & J0219+0120 & 1.4 & 37  &  6 (261)     & 30 \\
\hline
SHIZELS-3      & 2015.1.00026.S & 28 Jul.\ 2016 & J1058+0133 & J1058+0133 & J0948+0022 & 0.6 & 37  &  6 (261)     & 26 \\
SHIZELS-9      &                & 16 Jul.\ 2016 & J0238+1636 & J0238+1636 & J0217+0144 & 0.4 & 38  &  7 (344)     & 43 \\
SHIZELS-10     &                & 26 Jul.\ 2016 & J0238+1636 & J0238+1636 & J0217+0144 & 0.4 & 38  &  7 (344)     & 43 \\
SHIZELS-11     &                & 26 Jul.\ 2016 & J0238+1636 & J0238+1636 & J0217+0144 & 0.6 & 45  &  7 (344)     & 43 \\
               &                & 10 Aug.\ 2016 & J0006-0623 & J0006-0623 & J0217+0144 & 0.2 & 42  &       & 43 \\
               &                & 10 Aug.\ 2016 & J0238+1636 & J0238+1636 & J0217+0144 & 0.2 & 39  &       & 43 \\
SHIZELS-14     &                &  2 Aug.\ 2016 & J1058+0133 & J1058+0133 & J0948+0022 & 0.7 & 39  &  6 (261)     & 26 \\
\hline
\end{tabular}
\end{table*}

The High-$z$ Emission Line Survey \citep[HiZELS,][]{Sobral2012,
  Sobral2013, Sobral2015, Geach2008} was designed to study `normal’ star-forming
galaxies selected in narrow redshift slices at 0.4, 0.84, 1.47 and
2.23 via the identification of H$\alpha$ emission using near-infrared
(near-IR) narrow-band filter imaging in extragalactic survey fields including
the Subaru-{\it XMM} Deep Field / UKIDSS Ultra Deep Survey \citep[UDS, ][]{Lawrence2007} and the
Cosmological Evolution Survey \citep[COSMOS, ][]{Scoville2007} fields. 
The HiZELS survey detects thousands of emission-line
objects, samples the `typical' galaxy population \citep{Oteo2015, Cochrane2018}, following the so-called
`main sequence' for star-forming galaxies at $z\simeq$ 1.47 and 2.23
\citep{Gillman2019}, and probing below the knee of the H$\alpha$
luminosity function ($<\,L^*_{\rm H\alpha}$) at these redshifts
\citep[$L^*_{\rm H\alpha} = 10^{42.6} \rm erg\,s^{-1}$ at $z \simeq 1.47$ 
and $L^*_{\rm H\alpha} = 10^{42.9} \rm erg\,s^{-1}$ at $z\simeq 2.23$, ][]{Swinbank2012a, Cochrane2017}. 
Over thirty galaxies from HiZELS have also been
mapped in the follow-up near-IR using integral field unit (IFU) spectroscopy
(the `SHIZELS' sample) aided by adaptive optics (AO), with the
Spectrograph for INtegral Field Observations in the Near Infrared
(SINFONI) at the Very Large Telescope (VLT) or the Near-Infrared
Integral Field Spectrometer (NIFS) at Gemini-North
\citep{Swinbank2012a, Swinbank2012b, Molina2017, Gillman2019}. These
observations provide H$\alpha$ IFU imaging at $\sim$\,1\,kpc scales
for galaxies at $z \simeq $ 1.47 or 2.23.

In this work, we make use of the available AO-aided IFU data and the {\it
Hubble Space Telescope (HST)} F606W, F140W data, together with new
Atacama Large Millimeter/submillimeter Array (ALMA) observations
targeting the continuum emission at
submillimetre wavelengths at similar spatial resolution (synthesised beam of
$\sim0\farcs2-0\farcs3$ full width half maximum), to characterise the
spatial correlation between the the H$\alpha$/UV and dust emission of
high-redshift $z\simeq $ 1.47 and 2.23 galaxies (the `ALMA-SHiZELS' sample).
The combination of the AO-aided IFU H$\alpha$ data, {\it HST} data and the ALMA
observations provides a unique opportunity to characterise the spatial
correlation between the SFR tracers around the peak of cosmic star formation. 
We describe our observations, data reduction and analysis in Section 2. In Section
3, we present the results and discussion, and summarise in Section
4. Throughout this work, we assume a $\Lambda$CDM cosmology with
${\rm H_0}$\,=\,70\,${\rm km s^{-1} Mpc^{-1}}$,
${\rm \Omega_M} = 0.3$, ${\rm \Omega_\Lambda} = 0.7$, where
$0\farcs15$ corresponds to a physical scale of $\sim$1.3\,kpc at
both $z\simeq 1.47$ and 2.23.

\section{Observations and data reduction, and analysis}

\subsection{ALMA observations}

Nine SHiZELS galaxies were observed with ALMA during
Cycle 2, 3, 4 (projects 2012.1.00402.S, 2013.1.01188.S and
2015.1.00026.S; P.I.: E.\ Ibar) in Bands 6 or 7,
depending on whether the galaxies were at $z=2.23$ or 1.47,
respectively, corresponding to rest-frame $\sim$\,355$\mu$m.  The
observations were designed to detect continuum emission using four
spectral windows (SPWs), each covering an effective bandwidth of
1.875\,GHz at a spectral resolution of 15.6\,MHz. Observations were
taken under relatively good weather conditions with precipitable water
vapour (P.W.V.) ranging from 0.2\,mm to 3.9\,mm, and using 29 to 45
antennas (the earlier the cycle, the smaller the number) with 
the longest baselines spanning 1.0--1.5\,km.
The phase, bandpass and flux calibrators
for all observations used in this work are listed in
Table~\ref{tab:table1}.

Data reduction was carried out using the Common Astronomy Software
Applications ({\sc casa}) and using the ALMA pipeline up to the
production of calibrated $uv$ data products. Three imaging approaches
were performed using the task {\sc tclean}, exploring a Briggs
weighting ({\sc robust=0.5}), natural weighting, and $uv$ tapering
such that we created a synthesized beam of $\sim 1''$ (see
Figs~\ref{ALMAdet1}, \ref{ALMAdet2}).  In all cases, the signal is
interactively cleaned down to 2--3\,$\sigma$
(r.m.s.\,$\sim$\,25$\mu$Jy\,beam$^{-1}$) in regions with significant
emission at the source position. The astrometric accuracy of the ALMA
Band-7 image is around 1.5\,mas. Information about each target is
presented in Table~\ref{table2}.

Sources SHiZELS-7, SHiZELS-9, SHiZELS-11 and SHiZELS-14 were detected
by ALMA at a significance higher than 5$\sigma$ in at least one of the
three imaging approaches, while SHiZELS-21, SHiZELS-8, SHiZELS-10,
SHiZELS-2 and SHiZELS-3 remained undetected, regardless of the imaging
approach.

Within the field of view (FoV) of SHiZELS-7 and SHiZELS-10, we
identify two serendipitous detections. We denote them as SHiZELS7-ID2
and SHiZELS10-ID2 and their properties are listed in Appendix~\ref{APC}.

\begin{table*}
  \centering
  \caption{\label{table2} Properties of the SHiZELS galaxies presented in this work. ID and
     z$_{\rm H\alpha}$ and $M_\star$ are extracted from previous AO-aided IFU observations
    presented in \citet{Gillman2019}. $S_{\rm 355\mu m}$ and $R_{\rm dust}$ are the observed continuum
    flux density and the deconvolved half-light radius at 355\,$\mu$m (rest-frame).  $\ddagger$ Possible AGN.}
  \resizebox{\textwidth}{!}{
  \begin{tabular}{lccccccccccc} 
    \hline
    ID           & R.A.\ (J2000) & Dec. (J2000)  & $z_{\rm H\alpha}$ & $\log(L_{\rm H\alpha}/\rm erg\, s^{-1})$ & $S_{\rm 355\mu m}$ (mJy) & $\log(L_{\rm FIR}/{\rm L}_\odot)$  & $\log(M_*^{\rm MAGPHYS} /{\rm M}_\odot)$ & $\log(\rm SFR^{\rm MAGPHYS}/M_\odot \rm yr^{-1})$ & $A_v$  & $R^{\rm half-light}_{\rm dust}$ (kpc) & $R_{\rm dust}^{\rm uvfit}$ (kpc) \\
    \hline            
    SHiZELS-7    & 02:17:00.34 & $-$05:01:50.6  & 1.455 & 42.1 & 0.15    $\pm$ 0.03  & 10.5 $\pm$ 0.4     & 10.4 $\pm$ 0.2 & 1.6$\pm$0.3 & 0.8$\pm$0.5  & 3.7 $\pm$ 0.3 & 4.3 $\pm$ 0.4 \\
    SHiZELS-9    & 02:17:12.99 & $-$04:54:40.7  & 1.462 & 42.4 & 0.51    $\pm$ 0.06  & 11.2 $\pm$ 0.4     & 10.6 $\pm$ 0.1 & 1.9$\pm$0.2 & 1.0$\pm$0.3  & 4.9 $\pm$ 0.3 & 4.8 $\pm$ 0.4 \\
SHiZELS-11$\ddagger$& 02:18:21.23 &$-$05:02:48.9& 1.492 & 42.3 & 1.34    $\pm$ 0.14  & 11.6 $\pm$ 0.4     & 11.5 $\pm$ 0.1 & 1.7$\pm$0.1 & 0.8$\pm$0.1  & 5.2 $\pm$ 0.4 & 4.3 $\pm$ 0.5 \\
    SHiZELS-14   & 10:00:51.58 & $+$02:33:34.1  & 2.242 & 42.9 & 2.31    $\pm$ 0.23  & 12.6 $\pm$ 0.3     & 11.1 $\pm$ 0.1 & 2.6$\pm$0.2 & 1.5$\pm$0.2  & 4.7 $\pm$ 0.3 & 4.1 $\pm$ 0.4 \\
    SHiZELS-10   & 02:17:39.02 & $-$04:44:41.4  & 1.447 & 41.9 & < 0.11 (5$\sigma$)  & <11.3 (5$\sigma$)  & 10.1 $\pm$ 0.1 & 2.0$\pm$0.2 & 1.4$\pm$0.2  &    --         &    --         \\
    SHiZELS-8    & 02:18:20.96 & $-$05:19:07.5  & 1.460 & 42.2 & < 0.12 (5$\sigma$)  & <11.2 (5$\sigma$)  & 10.3 $\pm$ 0.1 & 1.9$\pm$0.2 & 1.4$\pm$0.4  &    --         &    --         \\
    SHiZELS-2    & 02:19:25.50 & $-$04:54:39.6  & 2.223 & 42.3 & < 0.12 (5$\sigma$)  & <11.9 (5$\sigma$)  & 9.9  $\pm$ 0.2 & 1.4$\pm$0.3 & 0.3$\pm$0.3  &    --         &    --         \\
    SHIZELS-3    & 10:00:27.69 & $+$02:14:30.6  & 2.225 & 42.4 & < 0.10 (5$\sigma$)  & <11.8 (5$\sigma$)  & 9.0  $\pm$ 0.1 & 0.9$\pm$0.2 & 0.2$\pm$0.1  &    --         &    --         \\
    SHIZELS-21   & 02:16:45.82 & $-$05:02:45.0  & 2.237 & 42.6 & < 0.12 (5$\sigma$)  & <11.8 (5$\sigma$)  & 9.8  $\pm$ 0.1 & 1.6$\pm$0.2 & 1.2$\pm$0.3  &    --         &    --         \\
    \hline
  \end{tabular}}
\end{table*}

\begin{figure*}	
\centering
\includegraphics[width=0.48\textwidth]{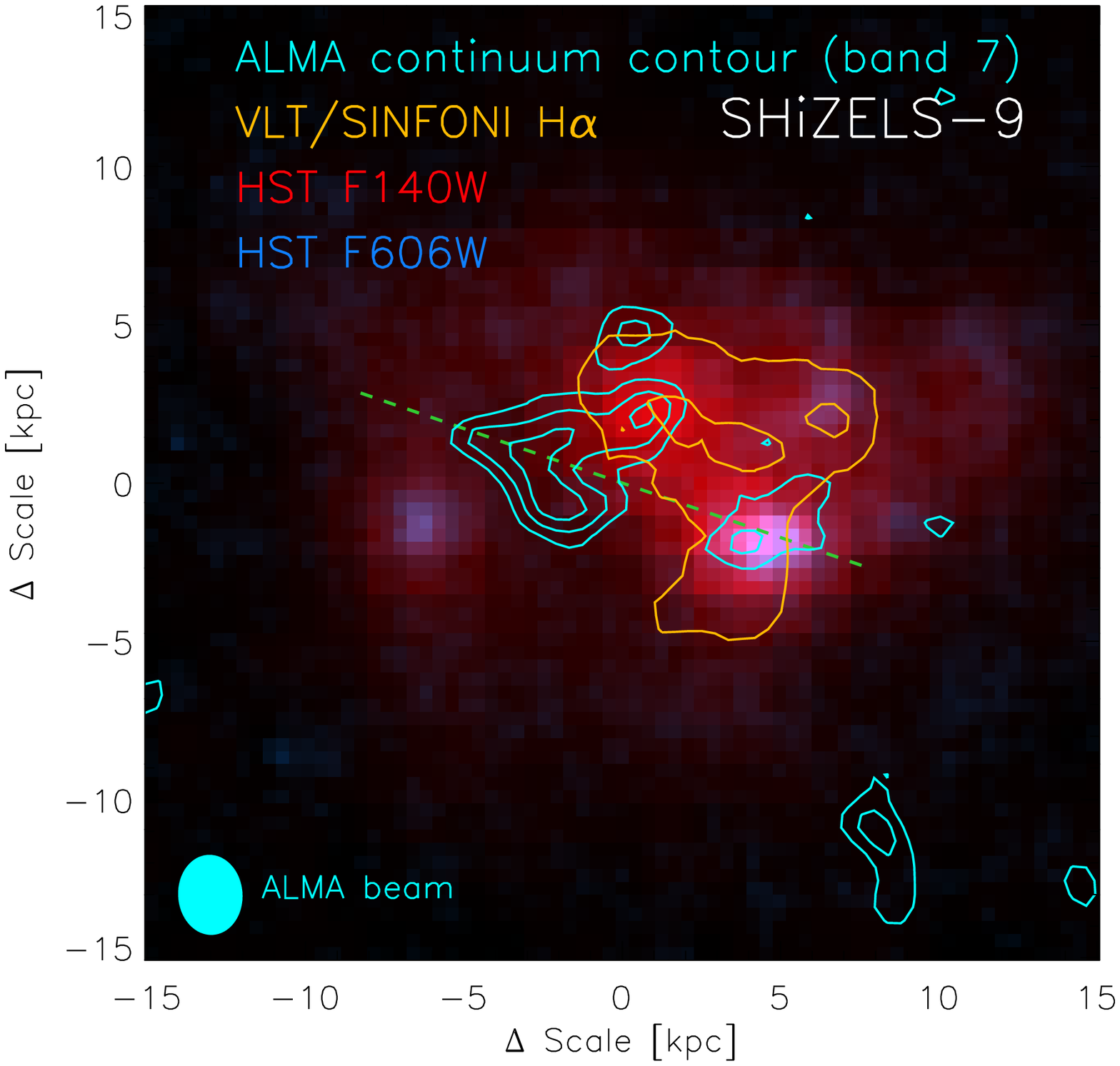}
\includegraphics[width=0.48\textwidth]{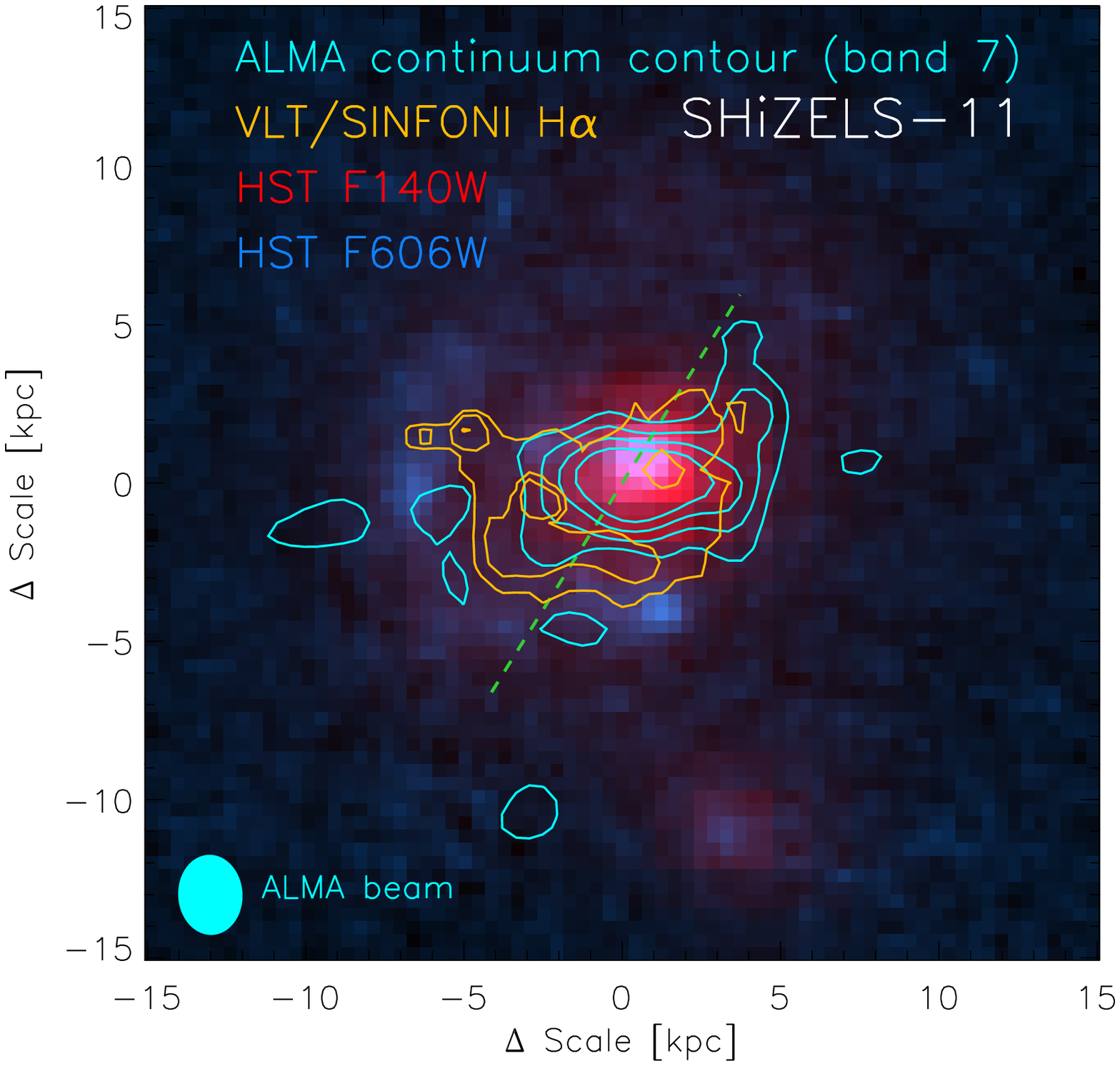}
\includegraphics[width=0.48\textwidth]{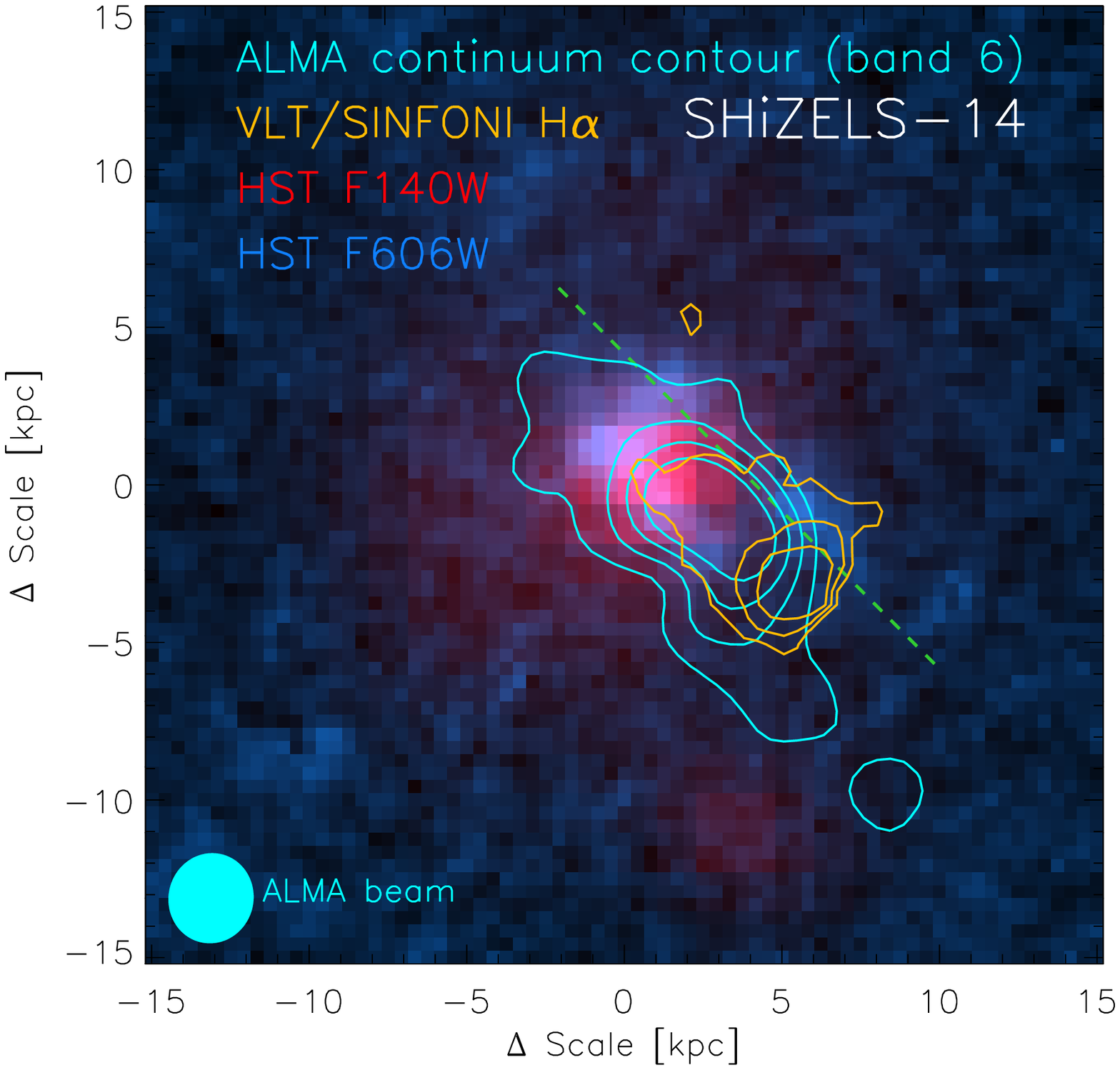}
\includegraphics[width=0.48\textwidth]{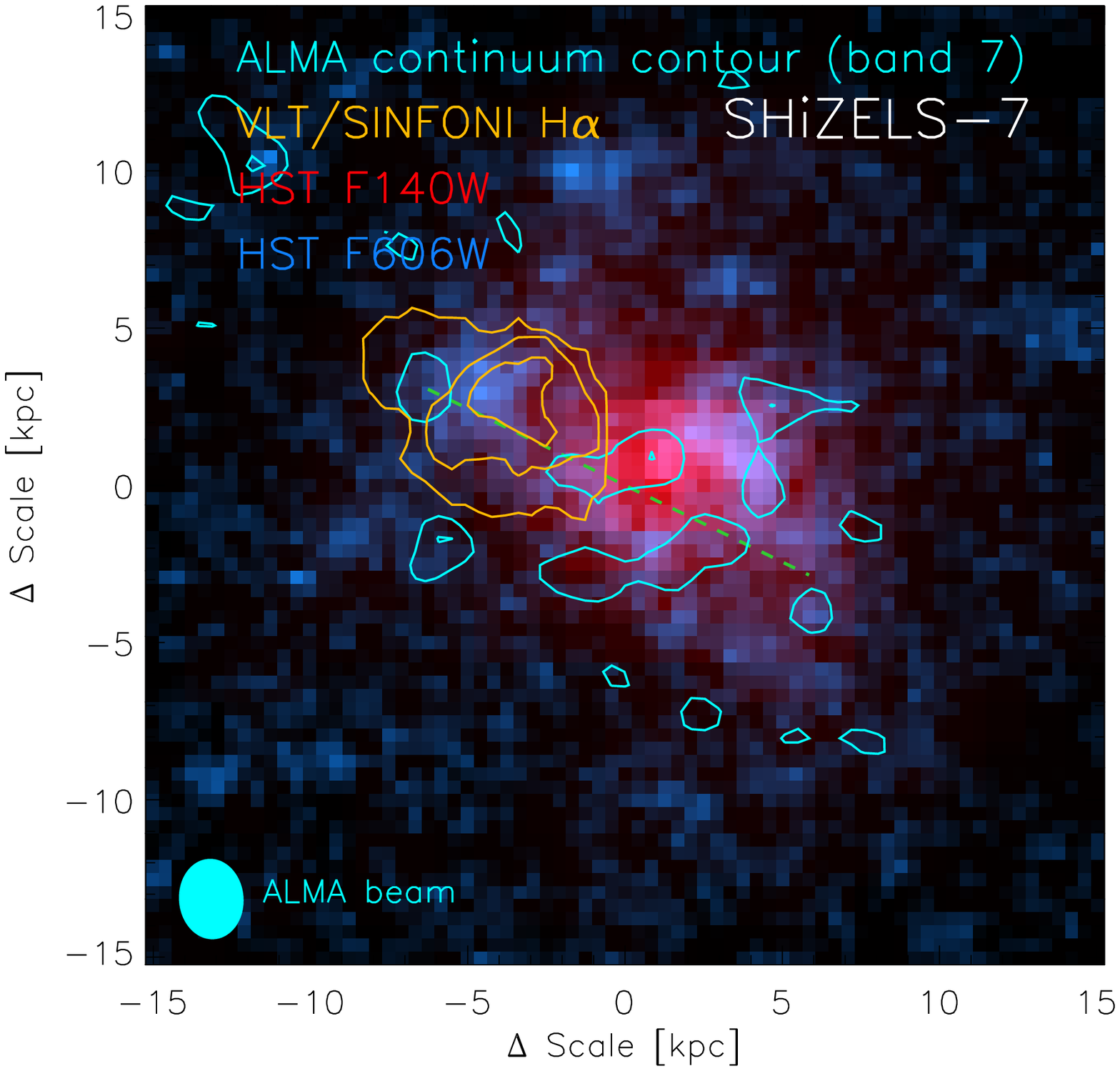}
\caption{The {\it HST} F606W (blue) and F140W (red) true-colour composite images, including ALMA
continuum contours (cyan) and H$\alpha$ contours (orange). Contour levels are shown at
2-, 3-, 4- and 5-$\sigma$, The ALMA synthesised beam is shown in the
lower left corner. The green dash lines show the direction of major kinematic axis \citep[$\rm PA_{vel}$, see the velocity map in][]{Gillman2019}.
The three different tracers of star formation are clearly
highlighting very different regions.}\label{colorimage}
\end{figure*}

\subsection{Global astrometry}
\label{astrometry-issue}

One of the main challenges when analysing multi-wavelength high-resolution
spatially-resolved observations is the global astrometric accuracy.
In particular, IFU observations with a small FoV (as obtained with
SINFONI, which has a FoV of about $3''\times 3''$) present a global
astrometric uncertainty which is larger than the resolution elements of
the observations, therefore they are difficult to anchor to other
observations at high-resolution. For this reason, we perform the
following astrometric corrections to analyse our data.

Firstly, we align the astrometry of the HiZELS narrow-band images by
the public {\it Gaia} DR2 catalogue.  The offset between the HiZELS
catalogue generated by SExtractor and the {\it Gaia} catalogue is
corrected to $\Delta {\rm R.A.} = 0\farcs004\pm 0\farcs13''$ and
$\Delta {\rm Dec.} = -0\farcs010 \pm 0\farcs10$.  We degrade the
resolution of the H$\alpha$ moment-0 SINFONI image down to $0\farcs7$
using a Gaussian kernel to get a similar resolution to the HiZELS
narrow-band images. Then we fit a Gaussian profile to this
low-resolution image in order to obtain the position of the peak,
identify the offset with respect to the narrow-band image, and use
this information to correct the astrometry of the IFU image.

As mentioned before, we have also used the available {\it HST} F606W and
F140W observations \footnote{{\it HST} proposal ID: 14719, P.I.:
  P.~Best}. The {\it HST} FoV is about $2'\times 2'$, which is too
small to apply a reliable astrometric correction using {\it Gaia}
DR2. Therefore, we correct the global astrometry of the {\it HST}
images using the HSC DR2 catalogue available in our fields
\citep{Aihara2019}. The 5$\sigma$ limiting magnitude within $2''$
diameter apertures of the HSC $i$-band catalogue is 26.7 AB mag, which is
similar to the {\it HST} images, and deep enough to have about 
200 high-S/N sources for the alignment. The astrometric accuracy
of HSC DR2 catalogue derives from the {\it Gaia} DR2 catalogue
\citep{Aihara2019}. We match the {\it HST} and HSC catalogues for the
offsets of R.A., Dec., and correct the astrometry of the {\it HST}
image. After our astrometric correction, the catalogue match between
{\it HST} and HSC catalogues is consistent with an offset of zero, with
systematic errors of $\sim 0\farcs04$.

In Fig.~\ref{colorimage}, we show the true-colour image composite
from {\it HST} F606W (in blue) and F140W (in red) images, including
ALMA continuum contours in levels of $2\sigma$, $3\sigma$, $4\sigma$,
$5\sigma$ for the four ALMA-detected galaxies. The galaxy centres
revealed by {\it HST} more or less coincide with the ALMA imaging.
The observations reveal a complex state for the
interstellar medium in SHiZELS galaxies at $z\simeq$ 1.47 and 2.23.
We show multi-wavelength postage stamps of the {\it HST}, H$\alpha$
\citep{Gillman2019} and our ALMA images for each of the galaxies
presented in this work in Figs~\ref{ALMAdet1} and \ref{ALMAdet2}.

\begin{figure}
\centering
\includegraphics[width=0.47\textwidth]{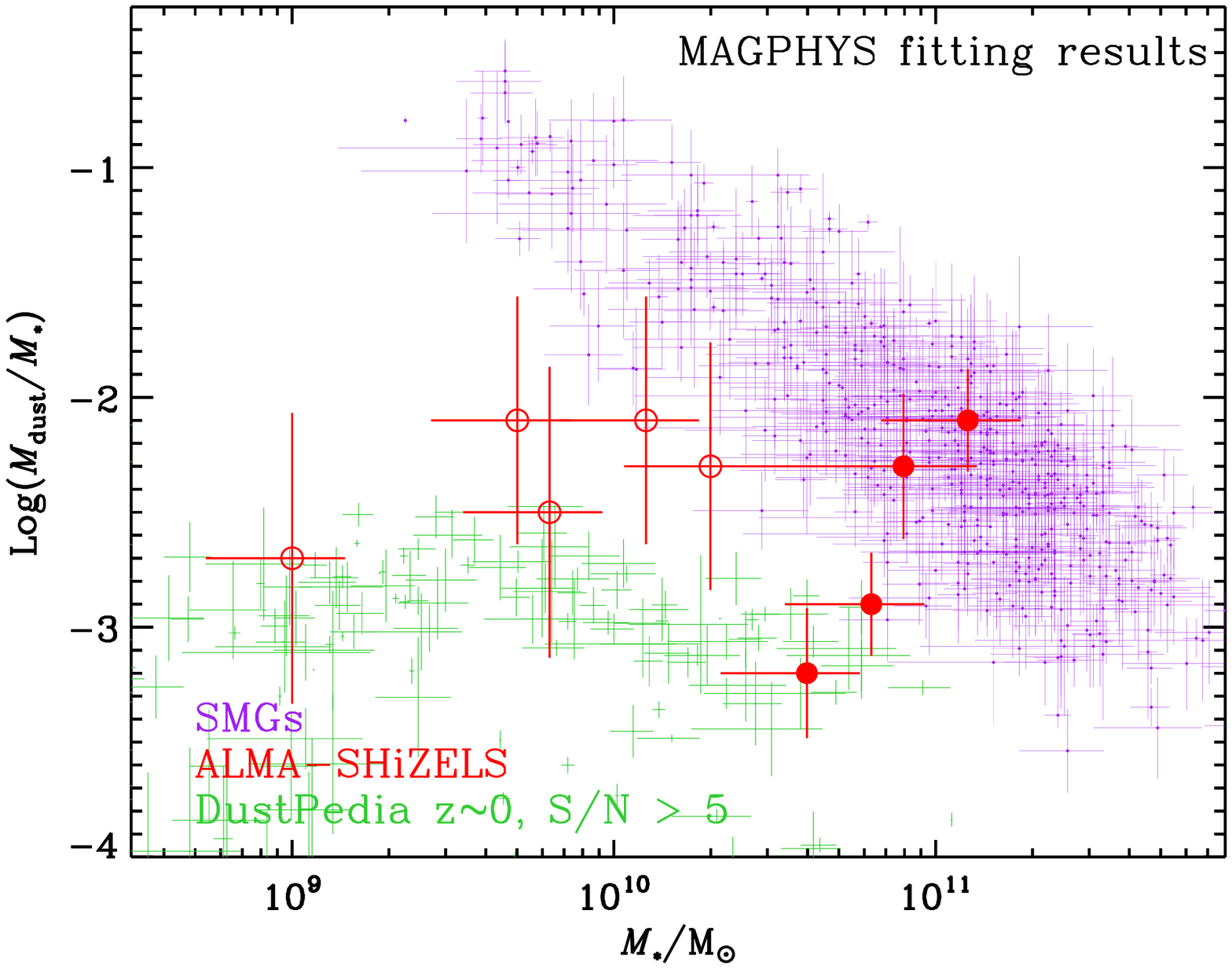}
\caption{The dust mass to stellar mass of our sample (red), 
the local galaxies from the Dustpedia project \citep[green, ][]{Clark2018},
and the SMGs from the ALMA SCUBA-2 UDS survey \citep[purple, ][]{Dudzeviciute2020}. 
The dust mass and the stellar mass of our sample and the comparison samples are derived 
from the SED fitting results by MAGPHYS.
The red filled circles are the ALMA detected targets while the red open circles are the ALMA non-detected targets.
The target with the stellar mass above $10^{11}M_\odot$ is SHiZELS-14, which is a ULIRG.
}\label{duststellarratio}
\end{figure}

\begin{figure}
\centering
\includegraphics[width=0.46\textwidth]{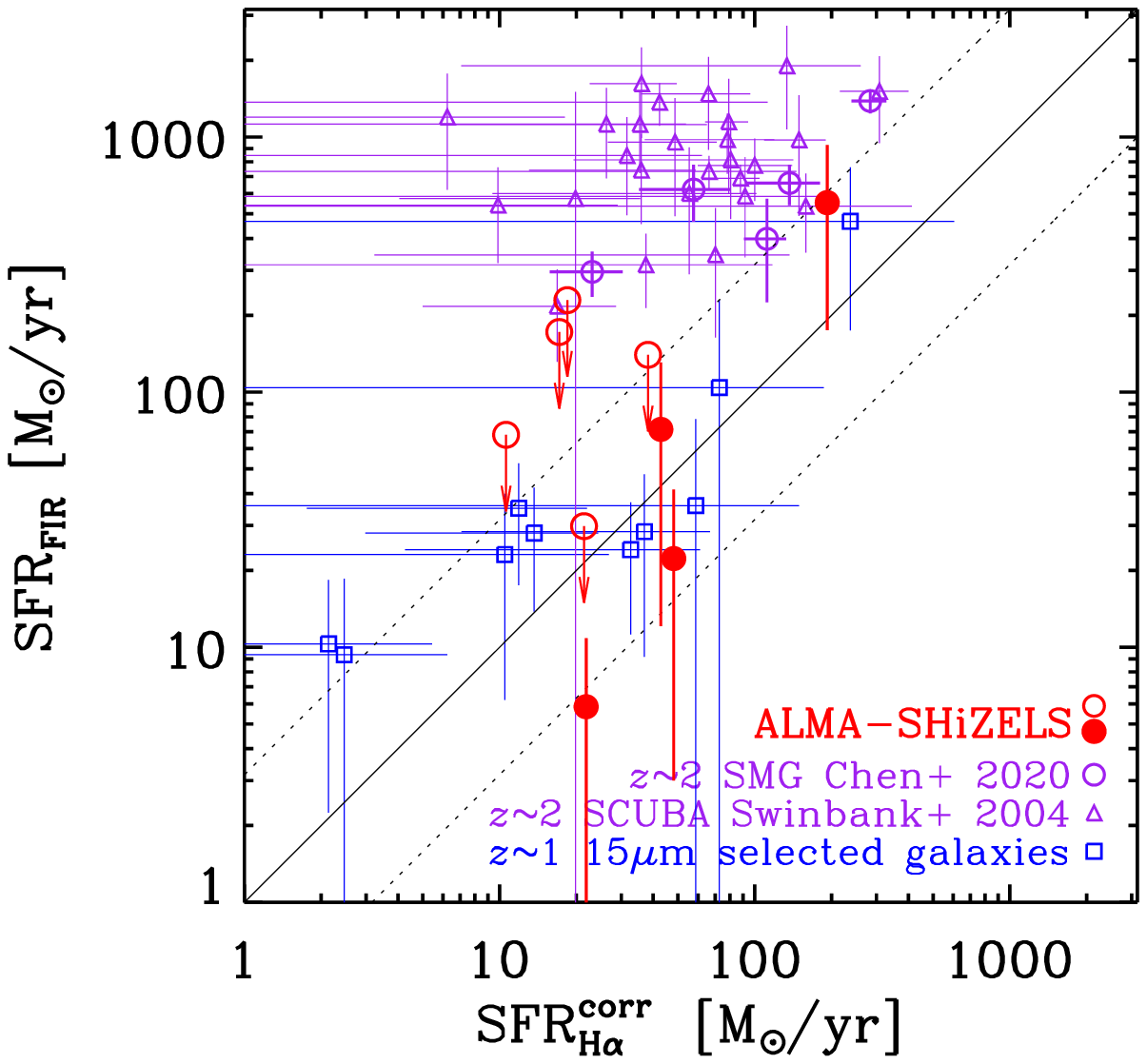}
\caption{The total SFRs derived from the extinction-corrected
H$\alpha$ (Eq.~3), and FIR luminosities for $z\sim 2.23$ and
$z\sim 1.47$ samples, respectively. We show the ALMA detected
targets as filled red circles and the ALMA non-detected targets
as open red circles. We compare our findings with
previously observed SMGs at $z\sim$2 from \citet{Swinbank2004} and
\citet{Chen2020}.  A sample of 15-$\mu$m-selected galaxies (at
$z\sim1$) presented by \citet{Franceschini2003} are also shown. 
The dotted lines show 0.5 dex above and below the solid line.}
\label{SFR_MASS}
\end{figure}

\begin{table*}
	\centering
	\scriptsize
	\caption{Gas phase metallicities derived from previous AO-aided IFU observations
	(following \citet{Curti2017}),
	gas to dust ratio (see \S~\ref{metalicity}),
	total ISM mass estimated following \citet{Scoville2016},
	the dust mass from gas-to-dust ratio  ($\log (M_{\rm dust}^{\rm GDR}$) and SED fitting  ($\log (M_{\rm dust}^{\rm SED}$)
	for the ALMA-SHiZELS galaxies presented in this work. $\ddagger$ Possible AGN. For the ALMA detected galaxy, the MAGPHYS fitting includes the new ALMA flux. }
	\label{tab3}
	\begin{tabular}{lcccccccccccc} 
		\hline
		ID         & [N\,{\sc ii}]/H$\alpha$ & 12+log(O/H)   & $\delta_{\rm GDR}$ & $\log(M_{\rm ISM}/{\rm M}_\odot)$  & $\log (M_{\rm dust}^{\rm GDR}/{\rm M}_{\odot})$& $\log (M_{\rm dust}^{\rm SED}/{\rm M}_{\odot})$ & $\log(M_{\rm dust}^{\rm MAGPHYS}/{\rm M}_{\odot})$ & $\log(M_{\rm dust}^{\rm MAGPHYS}/M_*)$ \\
		\hline                                                                   
		SHiZELS-7  &   0.43   $\pm$  0.05 & 8.79 $\pm$ 0.11  & 69 $\pm$ 24        &  9.65 $\pm$ 0.05             &7.8  $\pm$ 0.2     &7.5$\pm$ 0.2     & 7.4$\pm$0.2  & -3.2 $\pm$ 0.3\\
		SHiZELS-9  &   0.27   $\pm$  0.03 & 8.64 $\pm$ 0.11  & 97 $\pm$ 30        & 10.17 $\pm$ 0.04             &8.2  $\pm$ 0.1     &8.1$\pm$ 0.2     & 7.9$\pm$0.1  & -2.9 $\pm$ 0.2 \\
SHiZELS-11$\ddagger$ & 0.60   $\pm$  0.10 & 8.87 $\pm$ 0.12  & 57 $\pm$ 21        & 10.59 $\pm$ 0.06             &8.8  $\pm$ 0.2     &8.5$\pm$ 0.2     & 8.6$\pm$0.1  & -2.3 $\pm$ 0.3 \\
		SHiZELS-14 &   0.60   $\pm$ 0.05  & 8.87 $\pm$ 0.11  & 57 $\pm$ 21        & 11.19 $\pm$ 0.06             &9.4  $\pm$ 0.2     &8.8$\pm$ 0.1    & 9.0$\pm$0.1  & -2.1 $\pm$ 0.2 \\
		SHiZELS-10 &   0.13   $\pm$  0.04 & 8.47 $\pm$ 0.17  & 143 $\pm$ 37       &  $<9.50$                     &$<$7.3             &$<$7.4             & 8.0$\pm$0.5      & -2.1 $\pm$ 0.5  \\
		SHiZELS-8  &   $<$0.1             &$<$8.42           & $>$264             &   --                         &--                 &$<$7.4             & 8.0$\pm$0.5      & -2.3 $\pm$ 0.5  \\
		SHiZELS-2  &   0.12   $\pm$  0.01 & 8.45 $\pm$ 0.11  & 150 $\pm$ 38       &  $<9.92$                     &$<$7.7             &$<$7.6             & 7.3$\pm$0.6      & -2.5 $\pm$ 0.6  \\
		SHiZELS-3  &   0.03   $\pm$  0.01 & 8.13 $\pm$ 0.29  & 310 $\pm$ 52       &  $<9.84$                     &$<$7.3             &$<$7.5             & 6.3$\pm$0.6      & -2.7 $\pm$ 0.6  \\
		SHiZELS-21 &   0.23   $\pm$  0.04 & 8.60 $\pm$ 0.16  & 106 $\pm$ 32       &  $<9.92$                     &$<$7.9             &$<$7.6             & 7.6$\pm$0.5      & -2.1 $\pm$ 0.5  \\
		\hline                                                                
	\end{tabular}
\end{table*}

\subsection{MAGPHYS fitting with ALMA flux}
SEDs of the SHiZELS sample were previously fitted using MAGPHYS \citep{2008MNRAS.388.1595D} in \citet{Gillman2019}.
The ALMA detected fluxes of the ALMA-SHiZELS galaxies can help to constrain the FIR properties.
Therefore, we re-fit the SEDs of the four ALMA detected galaxies including the FIR flux we obtained in this work,
and list the main results in Table \ref{table2} and \ref{tab3}. For the ALMA non-detected 
galaxies, we list the dust mass given by MAGPHYS from fitting the optical-to-NIR SEDs.
The stellar mass and dust mass show reasonable consistencies from the different approach.

We show the dust-to-stellar mass ratio ($\log_{10}(M_{\rm dust}/M_*)$) in Fig. \ref{duststellarratio} 
and compare our results with the local galaxies from Dustpedia \citep{Clark2018, Casasola2020}.
The only ULIRG in our sample, SHiZELS-14, has a larger $\log_{10}(M_{\rm dust}/M_*) = -2.1\pm 0.2$ 
than the local galaxies, while this value is consistent with the typical value of $\log_{10}(M_{\rm dust}/M_*)$ 
in SMGs at redshift about 2 \citep{Calura2017}. The rest targets of our sample all have 
consistent or 1$\sigma$ higher $\log_{10}(M_{\rm dust}/M_*)$) values as the local galaxies, 
even for the targets with no ALMA detection.

\subsection{Global SFRs from H$\alpha$ and FIR emission}\label{Sect3.1}

To estimate the FIR luminosity (and SFR$_{\rm FIR}$) using the ALMA
observations (rest-frame continuum at $\sim$355\,$\mu$m), we assume a
FIR SED template based on previous stacking analyses for HiZELS
galaxies at $z \sim 1.47$ \citep{Ibar2013} and 2.23
\citep{Thomson2017}. Thanks to the rich multi-wavelength coverage, the
stellar masses of the HiZELS galaxies can be estimated reasonably well
\citep[see ][]{Sobral2014, Gillman2019}.  For the purpose of this
work, for the $z\sim1.47$ targets, we adopt the stacked SEDs derived
by \citet{Ibar2013} in the stellar mass bins of
$9.9<\log(M_* /M_\odot)<10.3$ and $10.3<\log(M_* /M_\odot)<11.8$. 
For galaxies at $z\sim2.23$, we consider the FIR template
presented in \citet{Thomson2017}.

We fit each FIR template SED following the modified-blackbody 
fitting method described by \citet{Beelen2006}, assuming a fixed
power-law index for the dust emissivity, $\beta = 1.8$. 
The fitting results show a dust temperature of about 25 $\pm$ 1 K 
for $z \simeq 1.47$ targets and 32 $\pm$ 2 K for the $z \simeq 2.23$ targets,
similar to previous measurements of luminous IR galaxies at high redshift \citep{Hwang2010, Oteo2017, Liang2019}.
To estimate FIR luminosities, we normalise the assumed SED to the observed ALMA flux densities. 
We derive dust masses following the method presented in \citet{Beelen2006}, 
finding values in the range of $10^{7.1} - 10^{8.9}$\,M$_\odot$.
From Table \ref{tab3} we can see that the dust masses derived from SED fitting and the
MAGPHYS are consistent. 
For the galaxies with no ALMA detection, we use global upper limits using 
$5\times$ the r.m.s.\ of the tapered image.

Uncertainties for the FIR properties come from the SED fitting, FIR
flux measured from ALMA results, and template assumption.  We expect
dust-temperature uncertainties for galaxy templates for
$M_* > 10^{10}$\,M$_\odot$ galaxies of about 5\,K \citep[see the
Fig. 5 in ][]{Ibar2013}, which leads to a systematic uncertainty in
$\log(L_{\rm FIR}/$L$_\odot)$ of about 0.3\,dex.  We run a Monte-Carlo
simulation to sample the dust temperature, assuming a Gaussian
distribution centred at $T_{\rm dust}$ from the FIR SED templates, and
a scatter $\sigma_{T_{\rm dust}} = 5$\,K. We also sample the ALMA flux
density (or flux density limit) from a Gaussian distribution centred
at the measured flux density with the observed scatter. Then we derive
$L_{\rm FIR}$ and $M_{\rm dust}$, and their r.m.s.\ scatter, such that
the scatter of $L_{\rm FIR}$ includes the uncertainty in
$T_{\rm dust}$ and ALMA flux density. Since we only have one detected FIR 
band for most sources, we cannot sensibly adopt a more complex model 
to understand the FIR properties.

The obscured SFR, as derived from the observed FIR emission, can be
estimated by \citet{Kennicutt2012}:
\begin{equation}
    \log {\rm SFR}_{\rm FIR}({M_{\odot}\,\rm yr^{-1}}) = \log L_{\rm FIR (8-1000\mu m)} ({\rm erg\,s^{-1}}) -  43.47,
\end{equation} 
assuming a Chabrier initial mass function \citep[IMF; ][]{Chabrier2003}.

\begin{figure}
\centering
\includegraphics[width=0.47\textwidth]{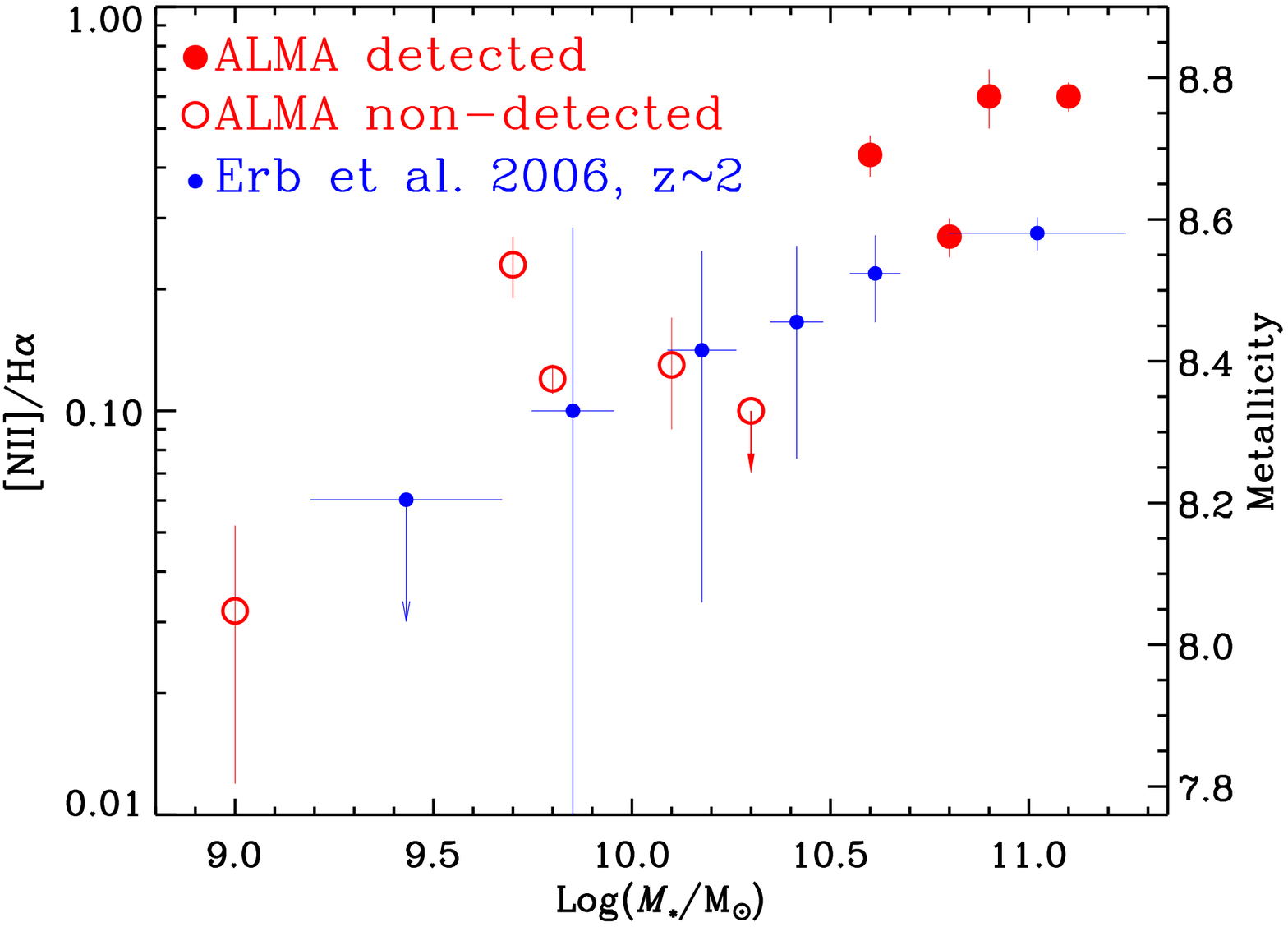}
\caption{The stellar mass versus [N\,{\sc ii}]/H$\alpha$ for our
  ALMA-SHiZELS galaxies. The ALMA-detected targets are shown in red
  while the non-detected targets are coloured blue. Since the
  metallicity could be derived from the [N\,{\sc ii}]/H$\alpha$ ratio \citep{Curti2017},
  we show the metallicity on the right-hand axis. We also show the
  mass-metallicity results at redshift $\sim$2 from
  \citet{2006ApJ...644..813E} in blue dots. Only the more massive
  galaxies with high metallicities are detected in continuum by
  ALMA.}\label{metamass}
\end{figure}

\begin{figure*}
\centering
\includegraphics[width=0.47\textwidth]{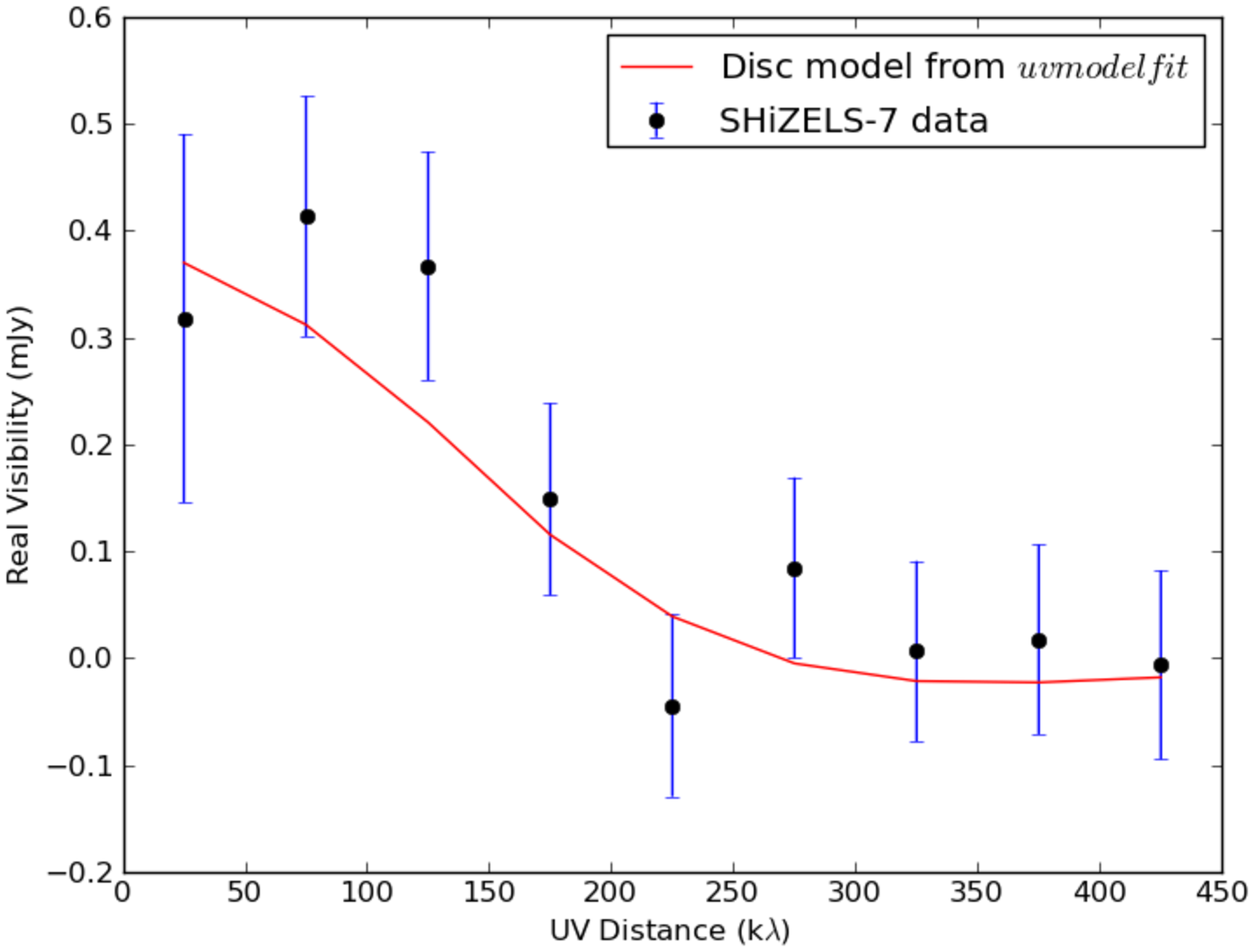}
\includegraphics[width=0.47\textwidth]{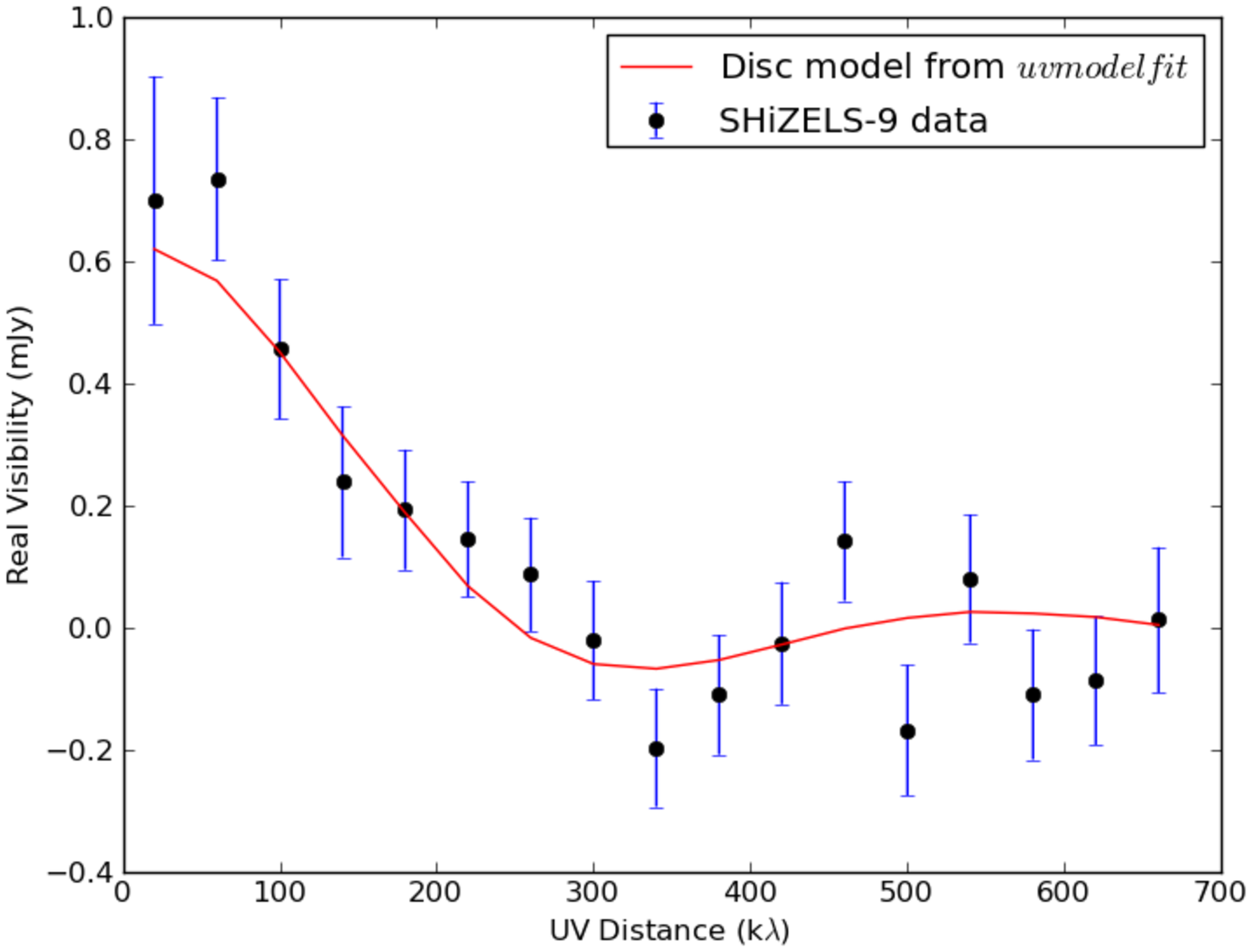}
\includegraphics[width=0.47\textwidth]{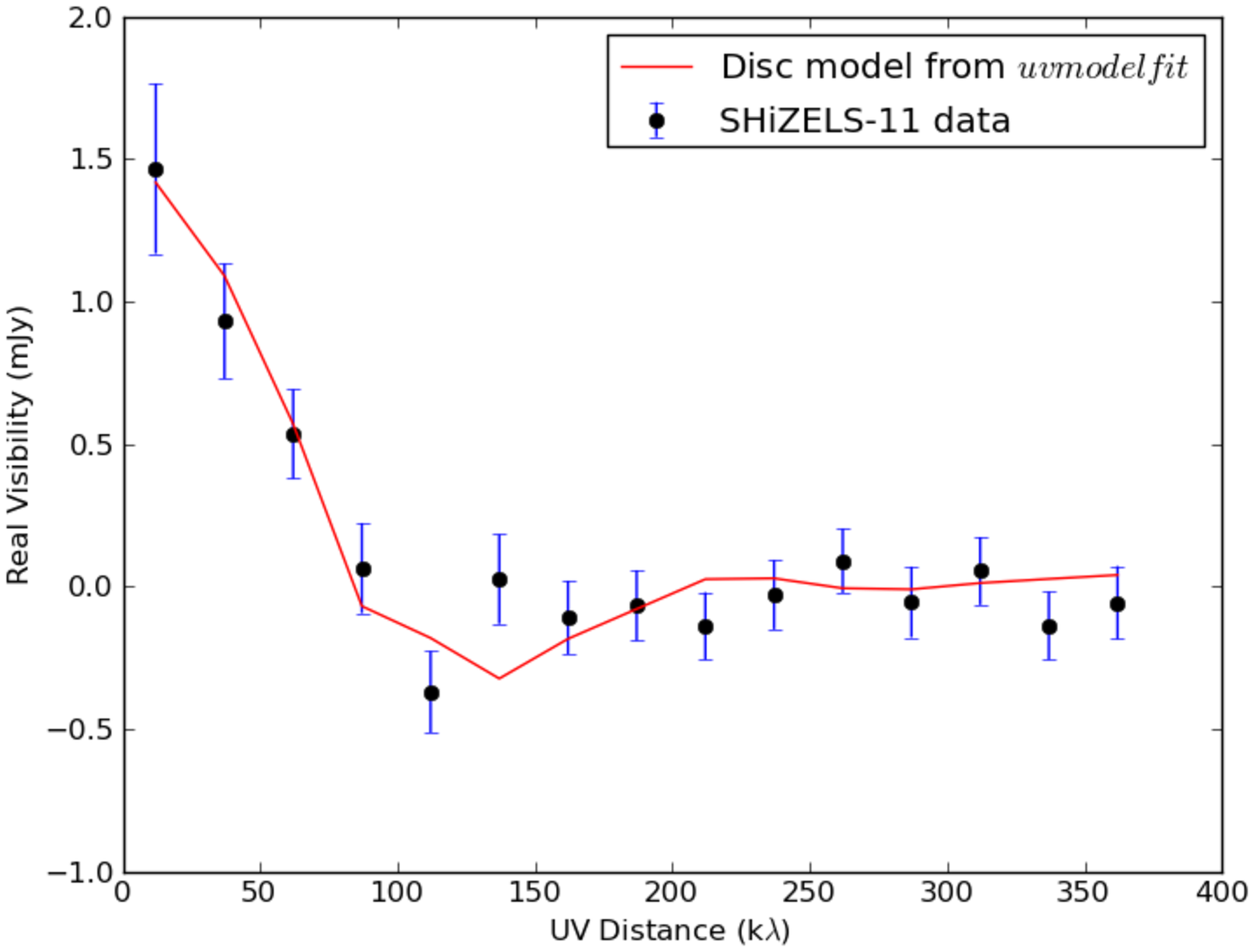}
\includegraphics[width=0.47\textwidth]{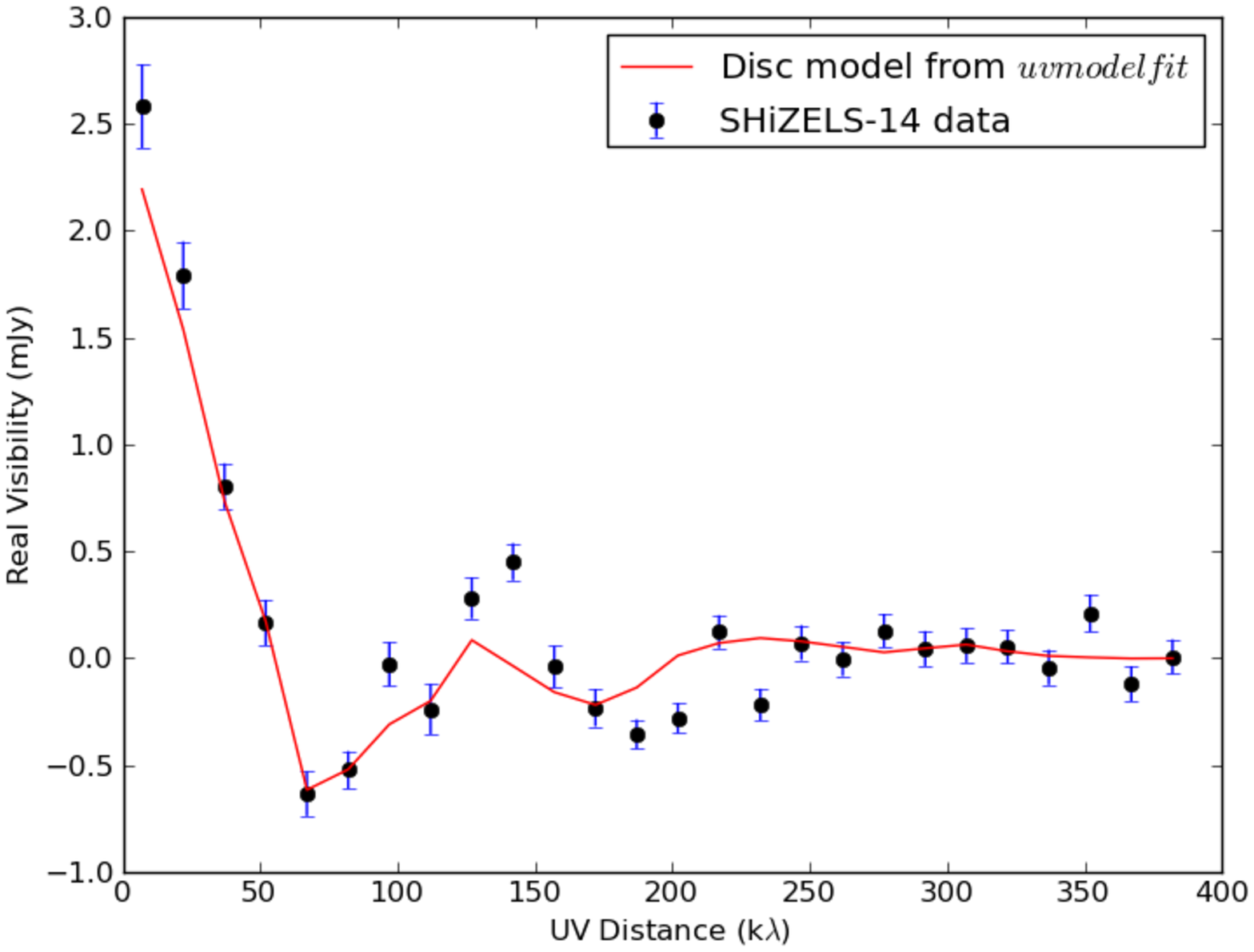}
\caption{The $uv$-real visibility diagrams of the ALMA detected targets. 
The black dots and blue error bar are obtained from the ALMA data while the red
lines are the fitting results by disk model in the CASA task {\sc uvmodelfit}.
}\label{uv-profile}
\end{figure*}

On the other hand, to correct the flux observed from the narrow-band
H$\alpha$ imaging for dust extinction, we assume a parametrisation of
the H$\alpha$ extinction as a function of stellar mass following
\citet{Garn2010}:
\begin{equation}
    A_{\rm H\alpha}(M_*) = -0.09 X^{3} + 0.11 X^{2} + 0.77 X + 0.91,
\end{equation}
where $X = \log_{10}( M_* / 10^{10} {\rm M}_\odot)$. 
This correlation between H$\alpha$ extinction and stellar mass
has also been confirmed by \citet{Sobral2012}, \citet{Koyama2019} and \citet{Qin2019}.
Using the stellar masses presented by in Table \ref{table2}, we derive the
extinction-corrected H$\alpha$ luminosity ($L^{\rm corr}_{\rm H\alpha}$) to obtain the SFR:
\begin{equation}
    \log {\rm SFR}_{\rm H\alpha}({{\rm M}_{\odot}\,\rm yr^{-1}}) = \log L^{\rm corr}_{\rm H\alpha} ({\rm erg\,s^{-1}}) -  41.27,
\end{equation}
assuming the same Chabrier IMF. 

A comparison between these two SFR estimates is shown in Fig.~\ref{SFR_MASS}. 

Previous studies of the SFR from FIR and H$\alpha$ show that SMGs at $z\sim 2$
have much larger SFR$_{\rm FIR}$ than SFR$_{\rm H\alpha}^{\rm corr}$ \citep{Swinbank2004, Chen2020}.

Due to the sensitive H$\alpha$ selection of our HiZELS parent sample, 
our sample comprises more typical `main sequence' galaxies, and is not 
limited to the most dusty starbursts. Only one of the ALMA-SHiZELS galaxies 
has FIR-derived SFR above 100 M$_\odot \,\rm yr^{-1}$. It is therefore not surprising 
that the SFRs derived from H$\alpha$ and FIR emission are more similar 
for our sample than for literature SMGs.

\subsection{The role of stellar mass and gas-phase metallicity}\label{metalicity}

We consider the gas-phase oxygen abundance as a proxy for metallicity,
and estimate it from the [N\,{\sc ii}]/H$\alpha$ emission line ratio
\citep{Pettini2004, Curti2017} from the near-IR IFU spectroscopy (see
Table~\ref{tab3}). The SHiZELS population has mainly Solar to
sub-Solar values (\citealt{Swinbank2012b,Molina2017}).  We show the
mass-metallicity relation of our ALMA-HiZELS sample in
Fig.~\ref{metamass} and identify the ALMA-detected and non-detected
targets.  We show the mass-metallicity relation at redshift $\sim 2$ from
\citet{2006ApJ...644..813E} as a comparison. In our observations, we
find that the lower the metal content or stellar mass, the weaker the
ALMA continuum emission. Indeed, all of the ALMA-detected galaxies
have [N\,{\sc ii}]/H$\alpha \gtrapprox 0.2$ or $M_*>10^{10.5}{\rm M}_\odot$, corresponding to
12+log(O/H)\,$>8.5$. The typical [N\,{\sc ii}]/H$\alpha$ emission line
ratio for galaxies at $z\sim2$ ranges from 0.03 to 0.3
\citep{Strom2017}, thus our ALMA-detected sample is indeed biased
towards massive and high-metallicity galaxies. As expected, 
dust continuum is more easily detected for a
galaxy with higher stellar mass or metallicity.

The ALMA detected targets have stellar mass values about one order of magnitude
higher than the ALMA non-detected targets. Since the SFR of the ALMA-SHiZELS 
sample is in the range of 10 to 100  M$_\odot$\,yr$^{-1}$ \citep[within one order of magnitude;][]{Gillman2019},
the ALMA detected targets in this work mainly have low specific SFR
compared to the rest of the SHiZELS sample.

Using the metallicity measurements, we can also derive the gas-to-dust
mass ratio ($\delta_{\rm GDR}$) following \citet{Magdis2012}:
$\log\delta_{\rm GDR} = (10.54\pm1.0) - (0.99\pm0.12)\times
[12+\log(\rm O/H)]$. 
We estimate ISM mass using the ALMA-derived 355\,$\mu$m flux, 
following \citet{Scoville2016} (Eq 16). 
Our derived dust masses are presented in Table \ref{tab3}. 
As shown there, these estimates are in good agreement with 
the values derived from SED fitting.

\subsection{Dust radius of the ALMA detected targets}
To estimate the dust radius, we fit the $uv$-real visibility of the 
ALMA detected targets by {\sc casa} task {\sc uvmodelfit}, and show the results 
in Fig. \ref{uv-profile}. We chose the $disc$ model in {\sc uvmodelfit} to fit 
the $uv$ profile. 
The circularised radius $R^{\rm uvfit}_{\rm dust} = \sqrt{R_{\rm maj}\, R_{\rm min}}$ of our targets are shown in Table \ref{table2}.
Since the FIR morphology is not regular, and thus the disk fitting results 
may not represent the flux distribution, we also measure the half-light radius ($R^{\rm half-light}_{\rm dust}$) 
of the ALMA images from the surface brightness distribution in the image plane
and show the results in Table \ref{table2}. The FIR surface brightness
profiles in real plane are also fitted by S\'ersic function and can be found in Section \ref{dustsurfacebrightness}.
The different approaches adopted for the measurements of the dust radii all show that the radii are 4-5 $\pm$ 0.5 kpc, 
larger than the typical size of the high-redshift SMGs \citep{Ikarashi2015, Simpson2015, Fujimoto2017}.
Dust emission from the four ALMA-detected galaxies is found to extend up to $\sim 2''$ ($\sim16$\,kpc, Fig. \ref{dust-profile}).

\section{Results and discussion}

\subsection{Individual galaxy properties}

Figs~\ref{colorimage} and \ref{ALMAdet1} and \ref{ALMAdet2} show the
{\it HST}, SINFONI and ALMA images of our SHiZELS sample.  For the
galaxies at redshift about 1.47 and 2.23, the {\it HST} F606W filter
corresponds to rest-frame 2400\,\AA\ and 1800\,\AA, therefore these
images are tracing mainly the detectable UV emission, i.e.\ the star
formation that is not obscured by the dust. On the other hand, the
observed {\it HST} F140W image reveals the rest-frame $r-$ or $g-$band
morphology (free of strong line emission), which is more sensitive to stellar mass.  Thus
Figs~\ref{colorimage}, \ref{ALMAdet1} and \ref{ALMAdet2} show proxies
for the stars (F140W), UV-traced star formation (F606W), H$\alpha$-traced star
formation (SINFONI), and dust mass distributions (ALMA). 
In this section, we describe properties of our
ALMA-detected targets individually\footnote{Most of our targets have
two versions of the H$\alpha$ maps \citep{Swinbank2012a, Gillman2019}. 
The main difference in the H$\alpha$ morphology is
caused by the different S/N criteria to create the H$\alpha$
maps. \citet{Gillman2019} built H$\alpha$ maps with high S/N spectra
while the results in \citet{Swinbank2012a} use all the available
spectra, which would then trace more extended H$\alpha$ emission.}.

We caution that morphological information derived from the rest-frame 
UV/optical can be potentially very misleading in sources with substantial 
dust extinction and on-going star formation and in particular it is 
difficult to trace the true stellar mass \citep[see ][]{Lang2019}.

\begin{itemize}
\item {\bf SHiZELS-7}: The H$\alpha$ kinematics reveal the presence of
  an extended disc-like rotating structure. Two H$\alpha$ star-forming
  clumps, separated by 4\,kpc, are identified by \citet{Swinbank2012a},
  although the fainter clump is not seen in \citet{Gillman2019}
  because of the different S/N criteria. The FIR continuum is spread
  over a diffuse structure. The two H$\alpha$ clumps are not likely to
  be affected by significant dust obscuration, so there might be no
  massive dust and gas clouds associated with the H$\alpha$ features. 
  The {\it HST} F140W and F606W images show that the least obscured regions in this galaxy
  show a compact stellar core and a extended UV morphology.

\item {\bf SHiZELS-9}: shows an extended rotation-dominated structure
  in H$\alpha$ with three bright clumps \citep[separated by $\sim$ 3\,kpc,][]{Swinbank2012a}. The FIR continuum shows a
  V-like structure in the central 3\,kpc, linking a bright stellar
  core with a fainter one. A third stellar core shows both UV and FIR
  emission.

\item {\bf SHiZELS-11}: This source was classified as a disk galaxy
  with a compact H$\alpha$ structure \citep{Swinbank2012a};
  although a newer analysis shows a marginally extended morphology
  \citep{Gillman2019}. SHiZELS-11 has an active galactic nucleus (AGN)
  identified in the X-rays by {\it XMM-Newton} and in the radio by the
  Very Large Array \citep[VLA ---][]{2006MNRAS.372..741S,
    2008ApJS..179..124U}. 
SHiZELS-11 is also detected by {\it Chandra} in X-UDS survey \citep{Kocevski2018}, with X-ray luminosity is $ L_{X} = 10^{43.3} $ erg $\rm s^{-1}$. 
The velocity dispersion of H$\alpha$ is about 90\,km\,s$^{-1}$
  \citep{Gillman2019}, much less than the typical velocity dispersion
  of type-I AGN, implying that the AGN in SHiZELS-11 must be
  obscured. Rotation is not clear from the H$\alpha$ dynamics, and the
  position of the maximum velocity dispersion is offset from the
  brightest H$\alpha$ pixel \citep{Gillman2019}. The {\it HST} F140W
  image shows a smaller source to the south, which may suggest a close
  merger or dust lane. The source has an apparently high metallicity
  and a steep metallicity gradient \citep{Swinbank2012a}, which might
  be due to contamination from an AGN. Both UV and FIR are bright in
  the galaxy centre, while the rest-frame optical is mainly coming from 
  a clumpy structure surrounding the centre. The ALMA continuum shows compact and extended
  emission, which does not overlap with the H$\alpha$.

\item {\bf SHiZELS-14}: This galaxy has been identified as a merger
  with three H$\alpha$ clumps separated by $\sim$ 5\,kpc
  \citep{Swinbank2012b}. The H$\alpha$ morphology in
  \citet{Gillman2019} is less clumpy, but still extends to a half-light
  radius of about 7\,kpc (Sersic model fitting results). The H$\alpha$
  kinematics show a velocity-dispersion-dominated system. The galaxy
  has the highest metallicity in the SHiZELS sample of
  \citet{Swinbank2012a}. The full extent of the H$\alpha$ emission is
  not traced by the rest-frame UV imaging. The ALMA continuum emission
  shows one dominant compact component at the centre, and more
  extended emission following a similar orientation as the F140W
  morphology. The FIR emission spreads up to $2''$,
  i.e.\ $\sim$16\,kpc at $z=2.23$. This target is comprehensively studied
  by Cochrane et al. (in preparation), including new Jansky VLA data.

\end{itemize}

\begin{figure*}
\centering
\includegraphics[width=0.98\textwidth]{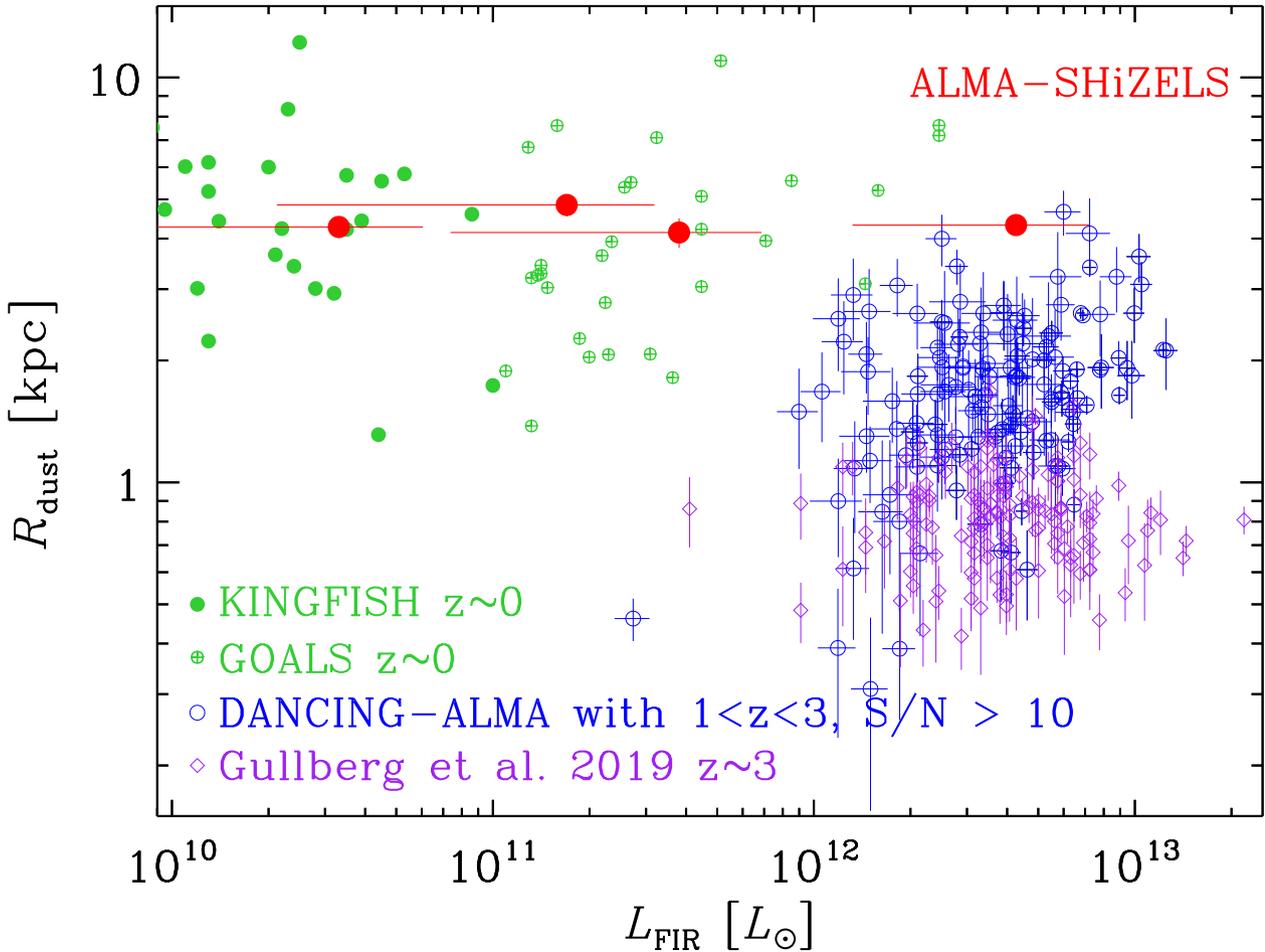}
\caption{Comparison between different surveys, local and high-$z$,
with previous spatially resolved FIR observations.
We show our ALMA-HiZELS galaxies (red dots), the galaxy sample from
KINGFISH \citep[green filled circles][]{Kennicutt2011},
GOALS \citep[green cross filled circles][]{Armus2009, Chu2017},
DANCING-ALMA with S/N of $L_{\rm FIR}$ higher than 5 \citep[cross filled blue circles][]{Fujimoto2017}
and the recent ALMA observation results of the SCUBA-2 bright galaxies \citep[purple diamonds]{Gullberg2019}.
The figure shows that our ALMA-SHiZELS are fainter than 
observations of typical sub-millimetre bright galaxies at similar redshifts, and also present larger sizes. 
The sizes of the SHiZELS galaxies compare better with those of low-redshift star-forming galaxies from KINGFISH and GOALS. 
}
\label{LIR_RE}
\end{figure*}

\begin{figure*}
\centering
\includegraphics[width=0.47\textwidth]{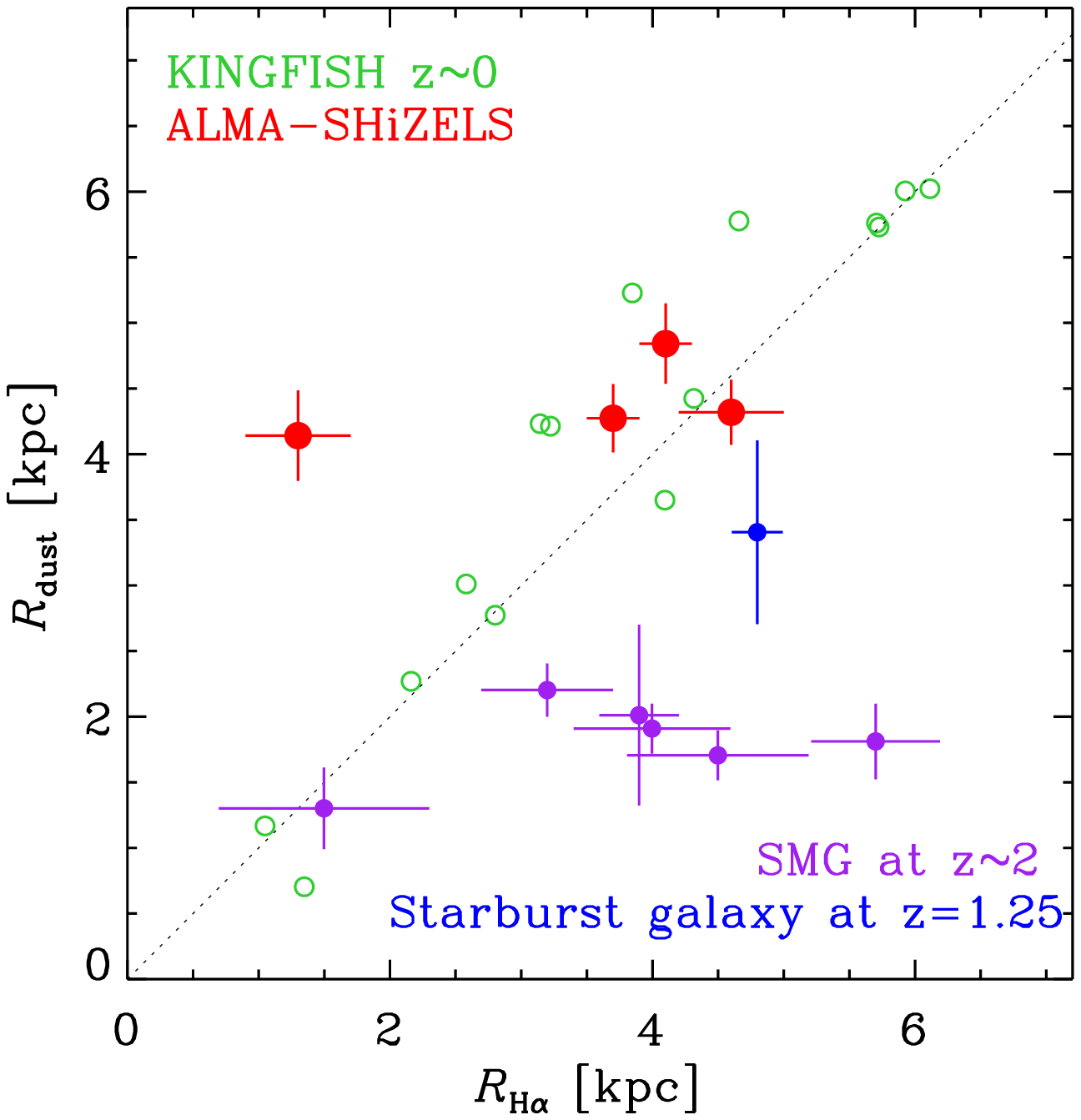}
\includegraphics[width=0.47\textwidth]{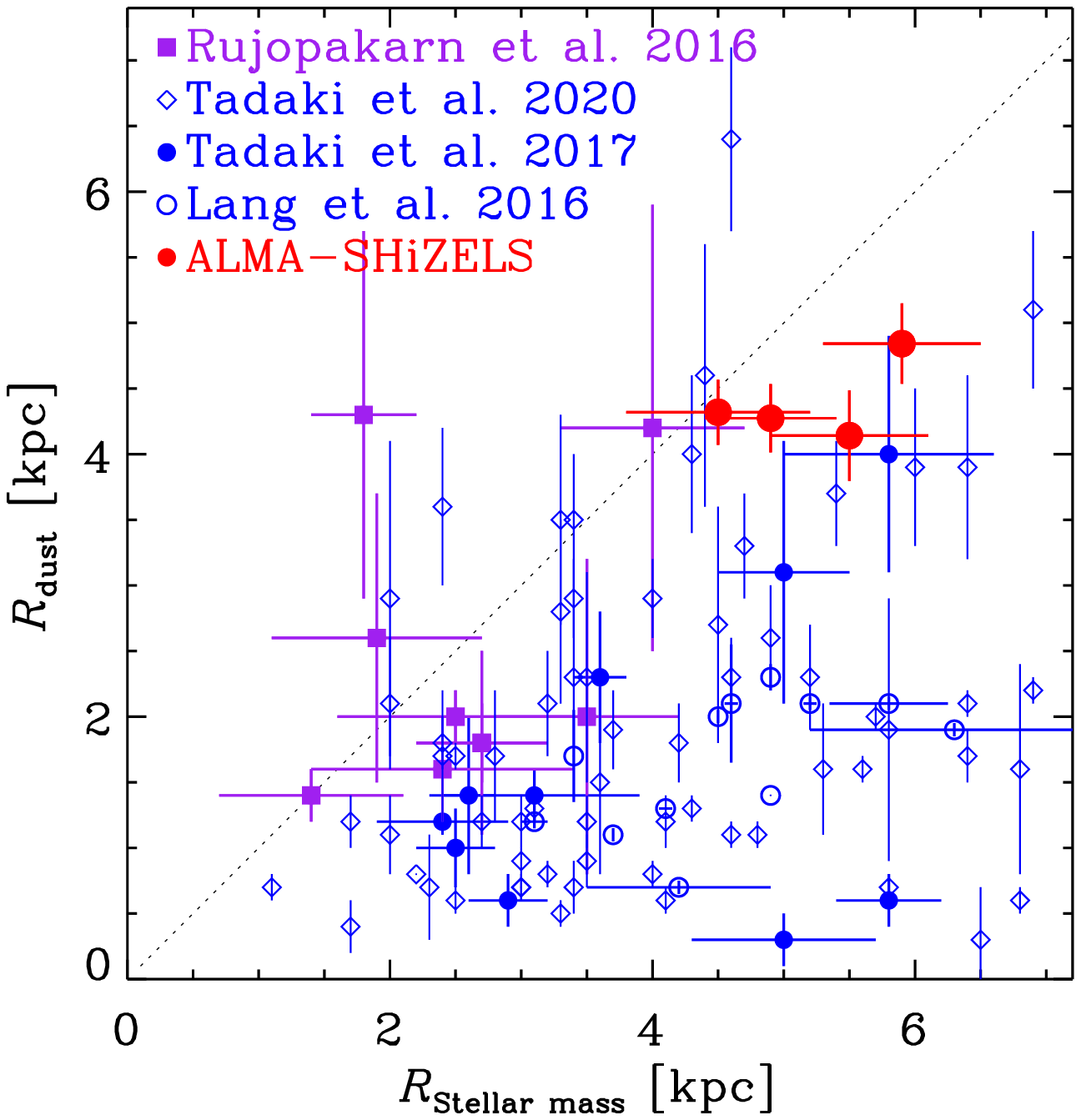}
\caption{
 {\bf Left panel: }Comparison of the half-light radii measured in the ALMA
  continuum imaging with respect to those measured in H$\alpha$
  \citep{Swinbank2012b, Molina2017}. We also overplot the {\it
    Herschel} 350-$\mu$m and narrow-band H$\alpha$ radii (obtained
  after convolving both to the same resolution) from the
  KINGFISH \citep{Kennicutt2011} and {\it Spitzer} Infrared Nearby
  Galaxies Survey
  (SINGS) projects \citep{Kennicutt2003, Dale2005, Dale2007}. Our
  high-redshift galaxy sample shows a broadly similar behaviour as is seen in
  these local galaxies, except SHiZELS-11 which shows a significantly
  larger radius in the FIR than in H$\alpha$. We also show the recent
  ALMA-resolved SMG sample at $z\sim2$ \citep{Chen2020} and one
  starburst galaxy (GOODSN-18574) at $z=$\,1.25 \citep{Nelson2019}.
{\bf Right panel:} The stellar mass size as derived from the {\it HST} 
F140W image compared to the FIR size as derived from the ALMA observations.
We show the H$\alpha$ selected main sequence star-forming galaxies at redshift about 2.2-2.5 
by \citet{Tadaki2017}, the star-forming galaxies presented 
by \citet{Rujopakarn2016}, the z$\sim$2 SMGs from \citet{Lang2019},
and the massive star-forming galaxy sample at z$\sim$2 from \citet{Tadaki2020}.
The dot line shows the one-to-one relation. 
Our targets are roughly the largest galaxies in both rest-frame optical band and FIR radius,
and may evolve into massive disk galaxies.
}
\label{HiZELS-Re}
\end{figure*}

\subsection{Spatially resolved properties}
\subsubsection{The dust emission from $z=1.47$ \& 2.23 galaxies}

Previous sub-millimetre studies of high-redshift galaxies at high spatial
resolution have revealed that clumps at $\sim$1\,kpc scales do not
particularly dominate the total flux emission, as is often seen in UV
or H$\alpha$ imaging \citep{Hodge2019}. For example, \citet{Hodge2016} used ALMA to
observe a sample of $z\sim2.5$ submillimeter galaxies (SMGs) at $0\farcs16$ resolution and
found that the observed 870$\mu$m continuum morphologies are
predominantly smooth and disc-like with typical radii of
$\sim 1.8$\,kpc. \citet{Gullberg2019} also shows that dust emission size from SMGs is 
about $\sim 1.5$\,kpc at $z \sim$\,1--4, with a larger sample.

Although these previous studies have revealed the internal properties
of the brightest SMGs, our focus is on the normal star forming galaxy population
\citep[i.e. galaxies selected by H$\alpha$ flux, with many having $\lesssim L_{\rm H\alpha}^*$][]{Swinbank2012a} at high redshift.
We find that our sample shows sub-millimetre continuum emission which is $\sim2\times$ more extended than
that seen in SMGs at $z\sim3$.
In the local universe, it is well known that ULIRGs present compact star-formation cores,
which may eventually lead to a compact stellar core. The large FIR radius of our ALMA-HiZELS
observations reveal the existence of a high-redshift galaxy population with extended
star-formation activity at $z=1.47$ or 2.23, possibly sustained in thick rotating disc-like
structures.

In Fig.~\ref{LIR_RE} we compare our results to previous spatially resolved FIR studies.
For the galaxies at $z=0$ and $L_{\rm FIR}<10^{11}L_\odot$,
the Key Insights on Nearby Galaxies: a Far Infrared Survey with {\it Herschel} 
project \citep[KINGFISH;][]{Kennicutt2011} surveys have found a typical radius 
of about $\sim$4 kpc with predominately disc-like morphologies \citep{Mosenkov2019}.
For isolated LIRGs taken from the Great Observatories 
All-sky LIRG Survey \citep[GOALS, ][]{Armus2009, Chu2017} we also find relatively large FIR sizes, similar to those from KINGFISH.
To have enough spatial resolution, we measure the GOALS galaxy FIR size based on the
{\it Herschel} PACS 160$\mu$m image. The difference between the FIR radius in 160$\mu$m and 350$\mu$m
is about 1.5 times or less \citep{Mosenkov2019}, which will not affect our results.
Some of the LIRGs in GOALS sample are galaxy pairs or merging galaxies. To avoid the contamination
from the neighbours, we only make use of the isolated galaxy sample here.

On the other hand, Demonstrating a New Census of Infrared Galaxies with ALMA \citep[DANCING-ALMA, ][]{Fujimoto2017} 
project presents $\sim$1000 galaxies ($0<z<6$ mainly ULIRGs) resolved
by ALMA at 1\,mm and finds a positive correlation between the radius
and $L_{\rm FIR}$ (in Fig.~\ref{LIR_RE} we only chose $1<z<3$, $S/N>10$ galaxies).
The recent work by \citet{Gullberg2019} shows compact FIR emission for $\sim$ 150 SMGs,
when the IR luminosity of the galaxy exceeds $10^{12}L_\odot$. 

Our ALMA-SHiZELS sample is different from the previous luminous high redshift
dusty populations as we are targeting a much fainter population. As shown in Fig. \ref{LIR_RE}
and Tab. \ref{table2}, our sample has typically lower IR luminosity, larger dust size 
than the high-redshift SMGs, which suggests that our survey are targeting the high-redshift 
star-forming galaxy population, rather than the high-redshift starburst galaxies.

The extension of the dust emission ($R_{\rm dust}$) could be roughly estimated
from a theoretical point of view, assuming $L_{\rm FIR} = 4\pi R_{\rm
  eff}^2\sigma T_{\rm mbb}^{4.32}$, where $\sigma$ is the
Stefan-Boltzmann constant, and the dust temperature ($T_{\rm mbb}$)
and $L_{\rm FIR}$ are
derived from SED templates (see \S~\ref{Sect3.1}; \citealt{Ma2015,
  Yan2016, Dudzeviciute2020}). This effective radius can be considered as the lower size
limit of the galaxy FIR emitters, hence the difference with respect to
the half-light radius could be understood as the spatial occupation of
dust structures below the resolution elements. Using this equation, we derive effective
radii for our SHiZELS galaxies of 0.3--1.1\,kpc, much smaller than the 
observed half-light radii of about 4.5 kpc, implying typical dust filling factors from 
5 to 10 for our sample. \citet{Hodge2019} have found that at 500\,pc 
resolution they can see tentative evidence of the spiral and bar structures of SMGs ($z=1.5-4.9$) 
at $\sim$250\,$\mu$m (rest-frame). Subject to surface brightness limitations, 
higher resolution ALMA imaging might reveal similarly complex structures in our galaxies.

\subsubsection{Spatial correlation between dust and ionised gas}

The observed morphologies of dust and H$\alpha$ emissions are well known 
to be correlated (at kpc scales) for samples of local star-forming galaxies, 
such as the KINGFISH project \citep{Kennicutt2011}. In the left panel of 
Fig. \ref{HiZELS-Re} we show that if we compare the half-light radii measured by 
ALMA with respect to the H$\alpha$ emission, we see good agreement for
sources SHiZELS-7, -9 and -14, suggesting spatial coexistence (at $\sim$ kpc scales)
of the dust and the ionised gas. In contrast, the source presenting an AGN,
SHiZELS-11, shows a significantly larger FIR radius than that seen in H$\alpha$.

We compare our results to the recent high resolution ALMA and H$\alpha$ observations of 
a sample of $z\sim 2$ SMGs \citep{Chen2020} and one starburst galaxy \citep{Nelson2019}
in the left-hand panel of Fig.~\ref{HiZELS-Re}. We find that our sample has a similar H$\alpha$ radius
to previously observed SMGs, suggesting extended star formation for these high-redshift 
galaxies. Despite this agreement, the dust emission from SMGs is typically more compact than 
our galaxies \citep{Gullberg2019}. 
The starburst galaxy in \citet{Nelson2019} has a dust radius higher than the SMGs, but
still lower than our sample.

Although the SHiZELS sample shows a larger dust radius than the SMGs and starburst galaxy,
most of the high-redshift galaxies in Fig. \ref{HiZELS-Re} have similar H$\alpha$ radius of about 4 kpc.
So the SMGs appear also to have an extended star formation region, as well as a compact dusty
core in the galaxy center (e.g., Fig. \ref{LIR_RE}). SMGs may obscure H$\alpha$ emission especially 
in the center kpc region, so the H$\alpha$ distribution in the SMG center might be more flatter, leading to a larger 
H$\alpha$ half-light radius. Since H$\alpha$ in SMGs can be bright and extended \citep{Swinbank2004, Swinbank2006}, 
while the dust morphology is compact \citep{Rujopakarn2019, Chen2020}, the large dust radii we observe in ALMA-SHiZELS 
sample may caused by the low SFR ($L_{\rm H\alpha} < L^*_{\rm H\alpha}$) of the SHiZELS sample,
rather than with the H$\alpha$-selection method.

The strong dust extinction in SMGs may also affect the H$\alpha$ emission 
such that the SFR derived from H$\alpha$ is lower than the SFR$_{\rm FIR}$ (Fig. \ref{SFR_MASS}), 
even though both SFR indicators may only reflect one aspect of the star formation 
and may still lower than the intrinsic SFR in galaxies. 

The SHiZELS sample was selected on H$\alpha$, down to a flux limit that traces below the knee of the luminosity function, 
and therefore the bulk of the sample will be less dusty than submm selected galaxies.
Our results show that the SHiZELS galaxies have very similar
dust and H$\alpha$ radii as the star-forming galaxies from KINGFISH at low redshift (Fig. \ref{LIR_RE} and \ref{HiZELS-Re}).
In the low-redshift universe, most of the star-forming galaxies are found to be spiral galaxies with low star-formation 
efficiency over an extended large disk \citep[e.g., ][]{Leroy2008, Cheng2018},
while ULIRGs usually have a more violent compact nuclear starburst in the galaxy center \citep{Solomon1997, Downes1998}. 
The similarity between the bulk of the ALMA-SHiZELS sample and the KINGFISH sample in Fig. \ref{LIR_RE} 
and \ref{HiZELS-Re} is in line with the two samples tracing more similar modes of main sequence star formation, 
compared with the extreme starburst modes witnessed in SMGs.

H$\alpha$ clumpy features are commonly found in high-redshift galaxies \citep{Genzel2006, Swinbank2012b}.
However, our ALMA continuum images do not show clear counterparts of the H$\alpha$ clumps, 
at least at $\sim$\,kpc scales (Fig. \ref{colorimage}).
It may be that the observed clumps are not tracing true star-forming clumps, 
but less dusty regions of the galaxy, from which the H$\alpha$ can escape \citep{Swinbank2004}.

\subsubsection{A comparison between the {\it HST} and ALMA morphologies}

The {\it HST} F140W images of our sample which trace the rest-frame optical emission
(Fig. \ref{ALMAdet1} and \ref{ALMAdet2}), reveal that only SHiZELS-9 has an apparently dual stellar core (the two cores have similar fluxes), 
which may be evidence of a major merger, or caused by a dust lane. SHiZELS-11 and -14 may both be undergoing or recently have undergone a minor merger event. 

Spatial offsets between the rest-frame optical and FIR observations 
have been found in low and high-redshift (U)LIRGs \citep{Charmandaris2004,  Hodge2016}.
In Fig.~\ref{colorimage} we see the spatial offsets and the overlap between the F140W and ALMA emission.
For SHiZELS-9, the two major (rest-frame) $V$-band cores present dust emission, while additional FIR 
emission comes from a region connecting to the third $V$-band core.
The $V$-band flux distribution does not always follow the H$\alpha$ or the dust morphology in our sample. 
A detailed study of spatially resolved dust extinction maps 
and the IRX-$\beta$ relation \citep[the FIR and UV luminosity ratio versus the UV SED slope, ][]{1999ApJ...521...64M} 
of our high-redshift star forming galaxies will be presented in a forthcoming paper.

The rest-frame UV images also show complex morphologies. Compared to the detected FIR emission from SHiZELS galaxies,
the UV morphologies are extended and clumpy. For the ALMA non-detected galaxies, SHiZELS-10, -2, 3, and -21, 
the rest-frame UV and optical band images show compact cores at their centres (Fig. \ref{ALMAdet2}),
evidencing compact UV star formation cores or regions of lower obscuration which is also found in the low-redshift low mass galaxies \citep{Cheng2020}. 
We see that $z\simeq 2.23$ galaxies display more compact rest-frame UV morphologies than the $z\simeq 1.47$ sample, which is 
consistent with previous studies \citep{Paulino2017}.

We show the half-light radii of the F140W emission in the right panel of Fig.~\ref{HiZELS-Re}.
The F140W radii (adopted from \citet{Gillman2019}) have typical effective radii of $\sim$\,4.5\,kpc, 
consistent with the typical size of the full HiZELS sample \citep{Stott2013}.
Using the stellar masses for our targets to predict half-light radius based on the mass-size relation \citep{Suess2019}, 
we find that the size revealed by the F140W imaging are consistent with the predicted radius within 1$\sigma$. 
We also present the galaxy sample from \citet{Rujopakarn2016} composed of star-forming galaxies at redshift $\sim 2$, 
observed by ALMA (870\,$\mu$m, 1.3\,mm) and VLA (5\,cm) at 0\farcs4, the H$\alpha$ selected main sequence star-forming galaxies from \citet{Tadaki2017}, 
the SMGs at redshift 2 observed by ALMA 870 $\mu$m from \citet{Lang2019}
, and the recent ALMA 870$\mu$m observation results of the massive star-forming galaxies at redshift 2 \citep{Tadaki2020}.
in the right panel of Fig.~\ref{HiZELS-Re}. \citet{Tadaki2017}'s sample of proposed elliptical progenitor galaxies has systematically 
smaller FIR size similar to other SMGs, suggesting the future formation of compact red cores. Our targets are roughly the largest galaxies in both rest-frame optical band and FIR radius.
The large FIR radius indicate the stellar mass is assembling at a larger radius, thus our targets would evolve into massive disk galaxies.

\section{Conclusion}

We present high-resolution ($0\farcs25$) ALMA continuum observations
(rest-frame $\sim355\mu$m) of nine star-forming galaxies at redshift 1.47
and 2.23, taken from the HiZELS survey. These galaxies have been observed with previous
AO-aided IFU H$\alpha$ spectroscopy and {\it HST} at similar resolution,
facilitating a spatial exploration of the star formation at high-redshift
at kpc scale resolution. Our sample comprises a population of star-forming
galaxies at redshift 1.47 and 2.23, which are mainly `main sequence'
galaxies, and have H$\alpha$ luminosities close to or below $L_{\rm H\alpha}^*$ at their redshift.

We detect four out of nine galaxies with ALMA. Their morphologies present extended faint structures,
out to 16\,kpc in diameter, much larger than the typical FIR size of sub-millimetre galaxies at high redshift. 
Our ALMA observations also reveal two serendipitous detections within the ALMA primary beams. 
Both of these are detected in previous optical and NIR surveys, located at different redshifts.
For the ALMA non-detected galaxies, we find they mainly have either lower stellar masses ($M_*<10^{10.5}{\rm M_\odot}$) 
and lower [N\,{\sc ii}]/H$\alpha$\,$<0.25$ ratios, corresponding to 12+log(O/H)\,$<$\,8.5.

The SFR derived from FIR and dust corrected H$\alpha$ are consistent with each
other within 1$\sigma$. However, the spatially resolved FIR and H$\alpha$ morphologies do not
show a similar distribution, implying a complex distribution of the ISM state in these galaxies.
At least at $\sim$kpc scales, the H$\alpha$ emission do not show a clear spatial correlation
with respect to the FIR emission. We find that the brightest
H$\alpha$ clumps, previously identified via AO-aided IFU spectroscopy,
are not significantly spatially correlated with the dust continuum emission which traces the bulk of the ISM, which appears smooth.

Our sample of SHiZELS galaxies have a typical rest-frame FIR size twice larger than the SMGs at high-redshift, 
while the H$\alpha$ emission size of our sample and SMGs are similar. The similarity between the 
extent of the dust continuum emission, and IR luminosities of our sample and the local star-forming disky galaxies 
is in line with our sample being drawn from the `normal' star-forming galaxy population at z=1.5-2.2.
The dust continuum emission in these systems is dominated by a cool extended component, 
while in more active galaxies such as SMGs, compact nuclear starbursts dominate the dust emission \citep{Gullberg2019}.

\section*{Acknowledgements}
This paper benefited from a number of thoughtful comments made by the anonymous referee.
C.C.\ appreciates useful comments from Dr.\ Wenda Zhang. 
This work is supported by the National Key R\&D Program of China grant 2017YFA0402704.
C.C. is supported by the National Natural Science Foundation of China (NSFC), No. 11803044, 11673028, and
J.H. is supported by the NSFC, No. 11933003. E.I.\ acknowledges partial support from FONDECYT
through grant N$^\circ$\,1171710. SG acknowledges the support of the Science and Technology Facilities Council through grant ST/N50404X/1 and ST/L00075X/1.
IRS acknowledges support from STFC (ST/T000244/1).
PNB is grateful for support from STFC through grant ST/R000972/1.
This work is sponsored in part by the Chinese Academy of Sciences (CAS), through a
grant to the CAS South America Center for Astronomy (CASSACA) in
Santiago, Chile.
This work was supported by the National Science Foundation of China (11721303, 11991052) and the National Key R\&D Program of China (2016YFA0400702).
A.E. acknowledges partial support from the Center of Excellence in Astrophysics and Associated Technologies (AFB-170002) and FONDECYT Regular Grant 1181663.
This paper makes use of the following
ALMA data: ADS/JAO.ALMA\#2012.1.00402.S, ADS/JAO.ALMA\#2013.1.01188.S,
ADS/JAO.ALMA\#2015.1.00026.S. ALMA is a partnership of ESO
(representing its member states), NSF (USA) and NINS (Japan), together
with NRC (Canada), MOST and ASIAA (Taiwan), and KASI (Republic of
Korea), in cooperation with the Republic of Chile. The Joint ALMA
Observatory is operated by ESO, AUI/NRAO and NAOJ. A work based on
observations collected at the European Organisation for Astronomical
Research in the Southern Hemisphere under ESO programmes 082.B-0300(A)
and 092.A-0090(A). This research is based on observations made with
the NASA/ESA Hubble Space Telescope obtained from the Space Telescope
Science Institute, which is operated by the Association of
Universities for Research in Astronomy, Inc., under NASA contract NAS
5--26555. These observations are associated with program 14719.

{\it Data availability:} The data underlying this article
are available in http://almascience.nrao.edu/aq/ and
https://archive.stsci.edu/index.html, and can be accessed 
with ALMA project ID: 2012.1.00402.S; 2013.1.01188.S;
2015.1.00026.S and HST proposal ID: 14719

\appendix

\section{Stamp images of ALMA-HiZELS targets}
\begin{figure*}
\centering
\includegraphics[width=1\textwidth]{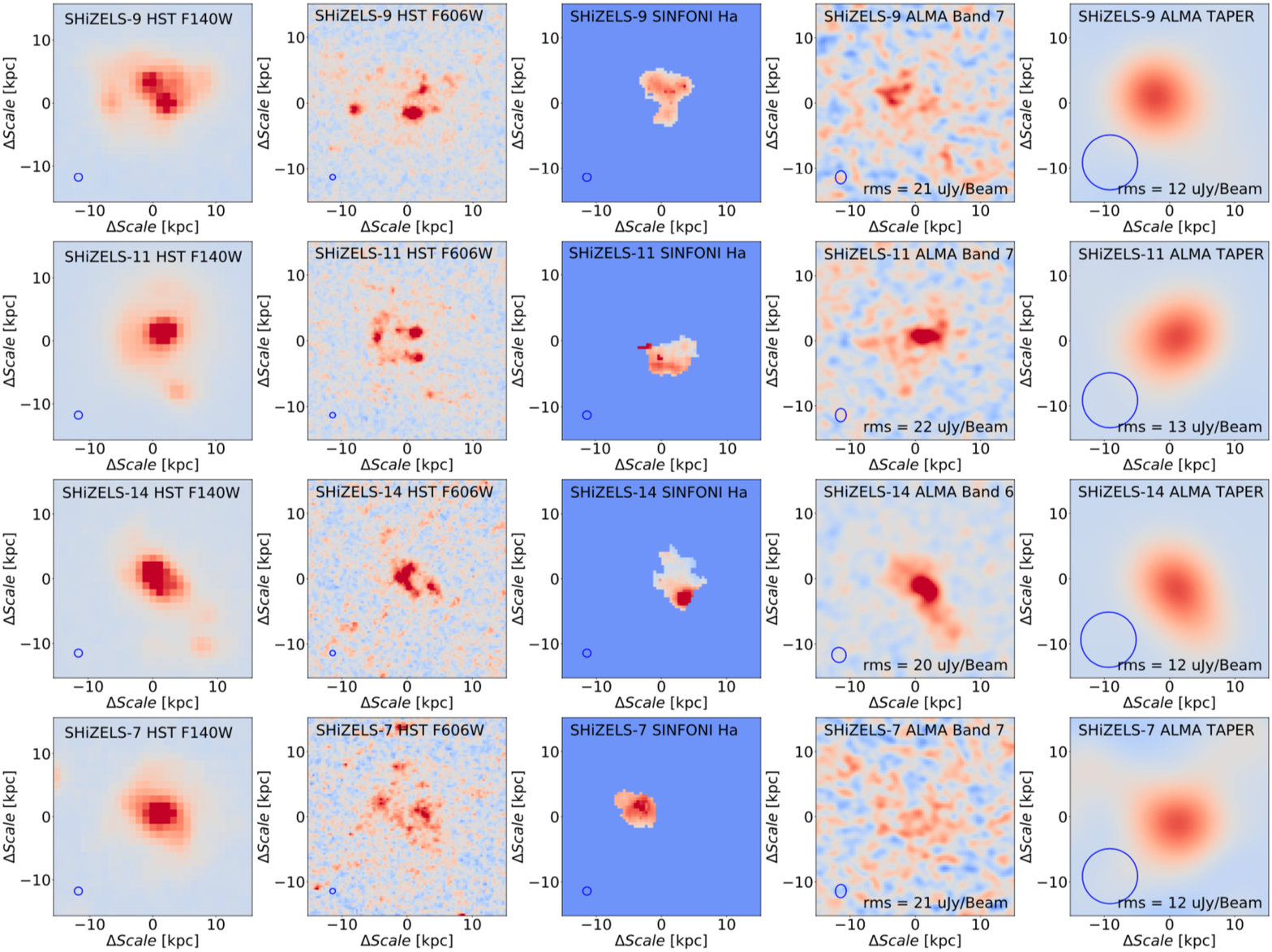}
\caption{Postage stamp (30 kpc $\times$ 30 kpc) images for the SHiZELS galaxies presented in
  this work. From left to right images: {\it HST} F140W, {\it HST} F606W,
  VLT/SINFONI H$\alpha$ moment-0, Briggs weighting ({\sc robust=0.5})
  ALMA continuum at $\sim 355\mu$m rest-frame, and tapered ALMA image
  (synthesised beam of $\sim 1''$) to highlight extended emission.
  The image shows the complexity of the different phases of the ISM in these
  high-redshift galaxies}
\label{ALMAdet1}
\end{figure*}

\begin{figure*}
\centering
\includegraphics[width=1\textwidth]{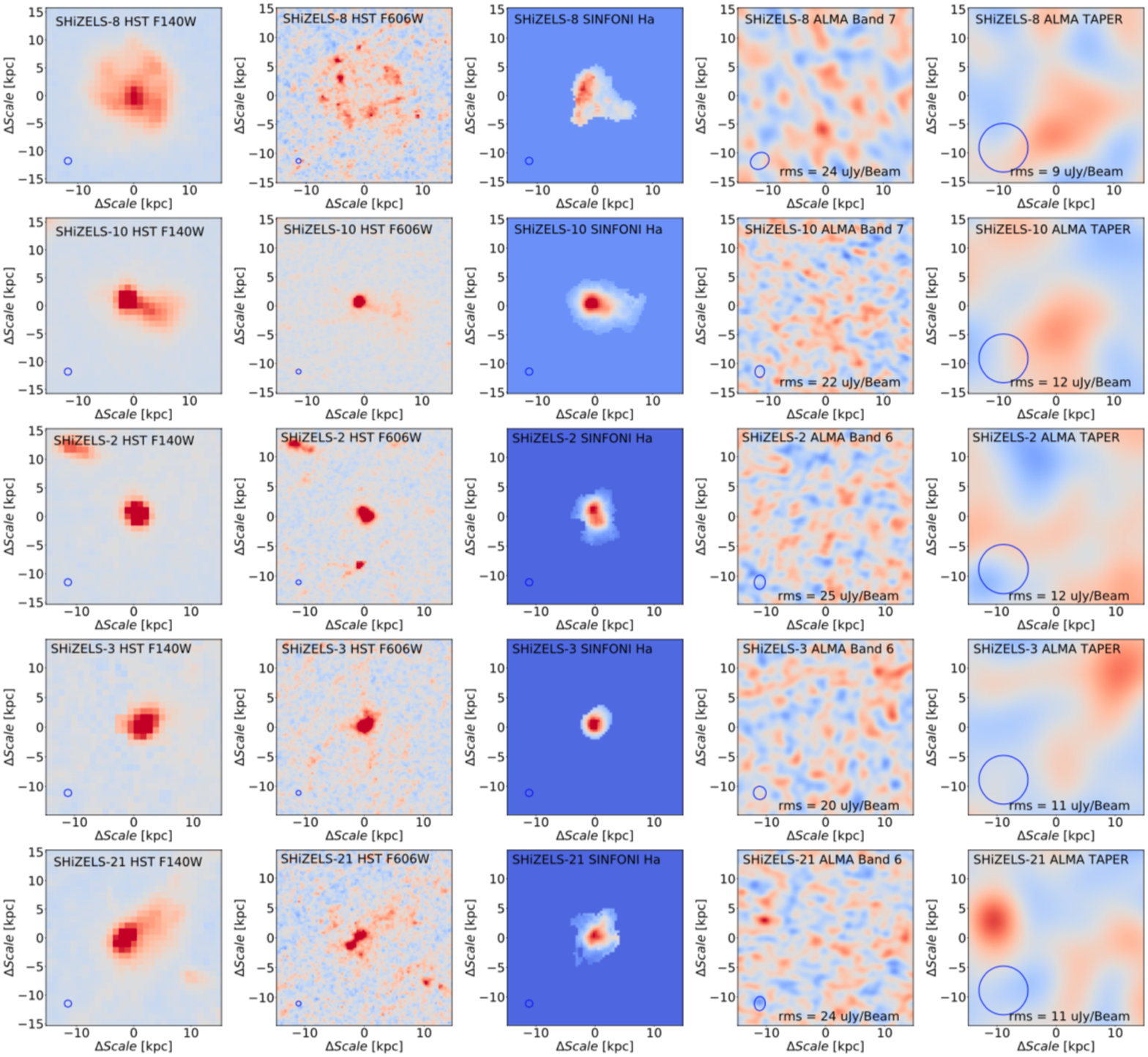}
\caption{Same as Fig.~\ref{ALMAdet1}, but for the undetected ALMA continuum sources.}\label{ALMAdet2}
\end{figure*}

We present the {\it HST}, VLT/SINFONI, and ALMA images in Fig. \ref{ALMAdet1} and Fig. \ref{ALMAdet2}.

\section{Dust surface brightness profiles}\label{dustsurfacebrightness}

\begin{figure*}
\centering
\includegraphics[width=0.46\textwidth]{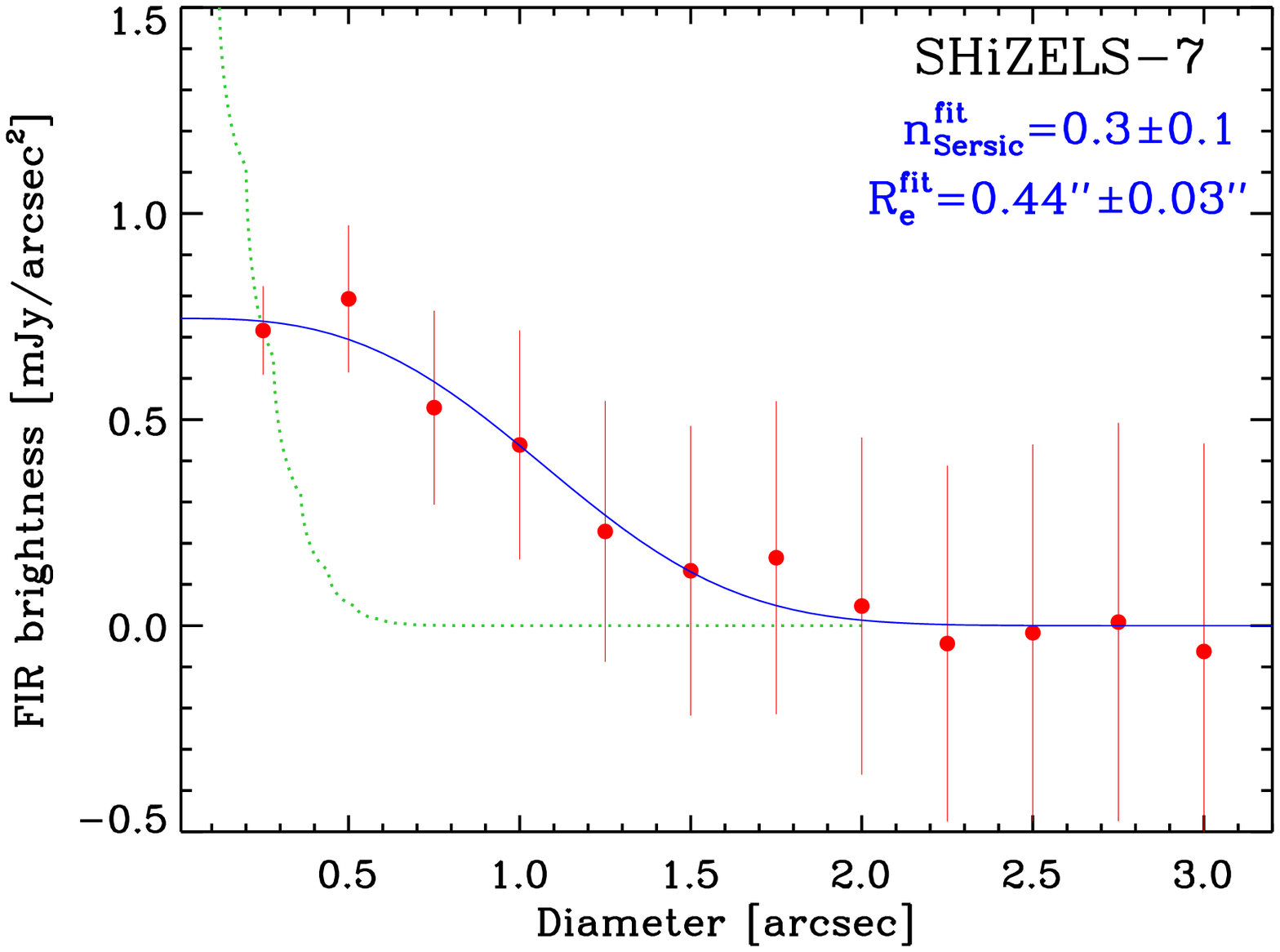}
\includegraphics[width=0.46\textwidth]{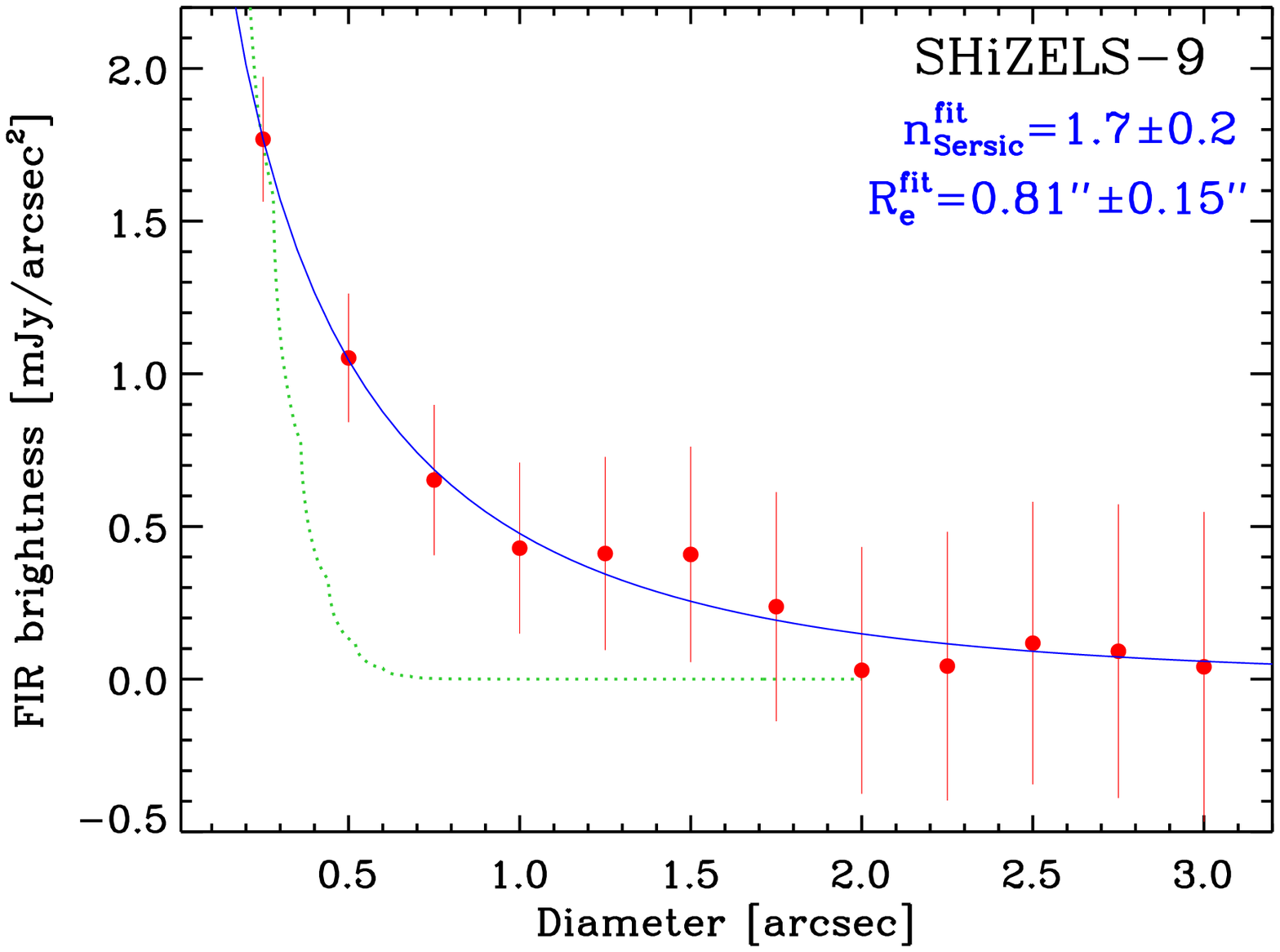}
\includegraphics[width=0.46\textwidth]{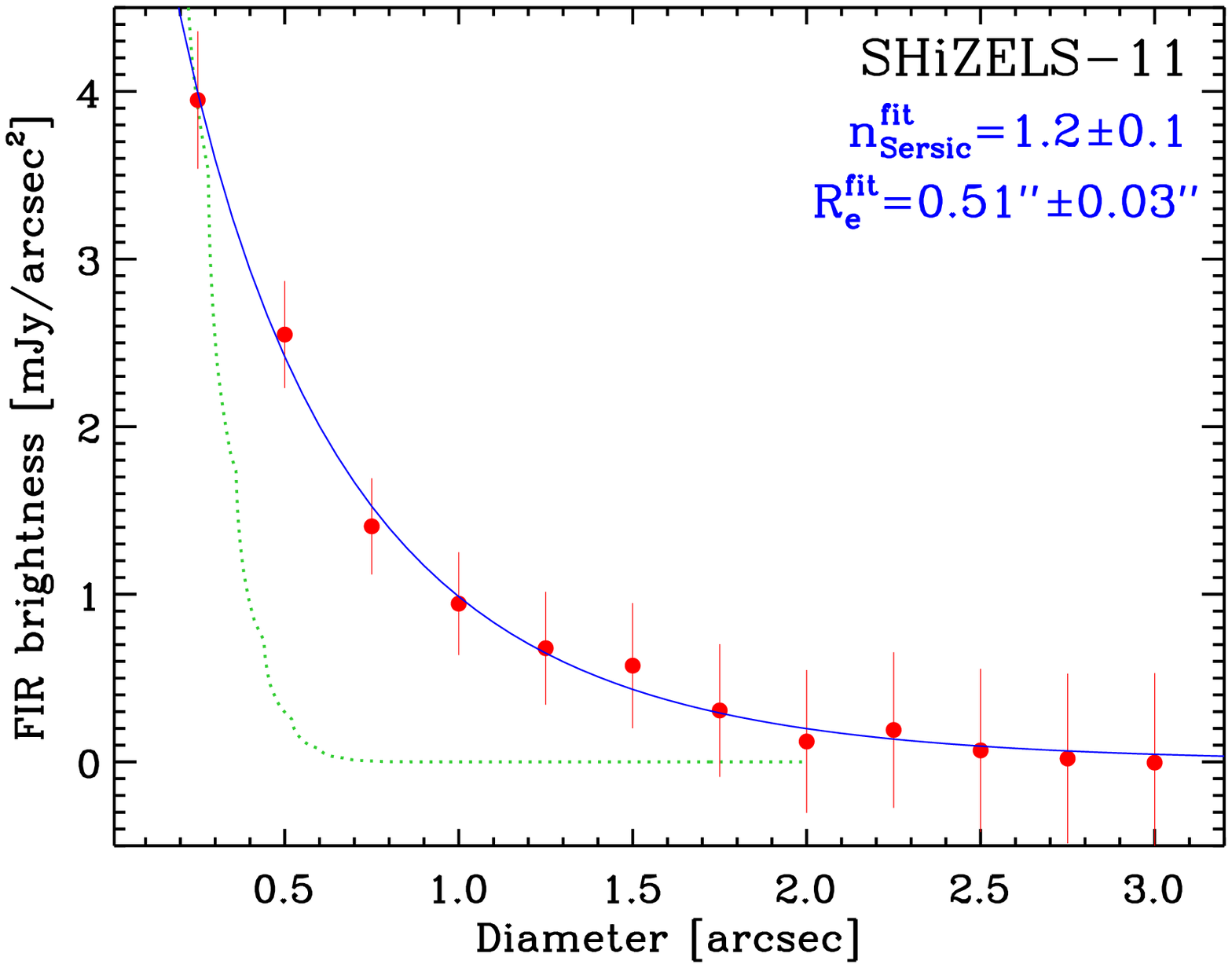}
\includegraphics[width=0.46\textwidth]{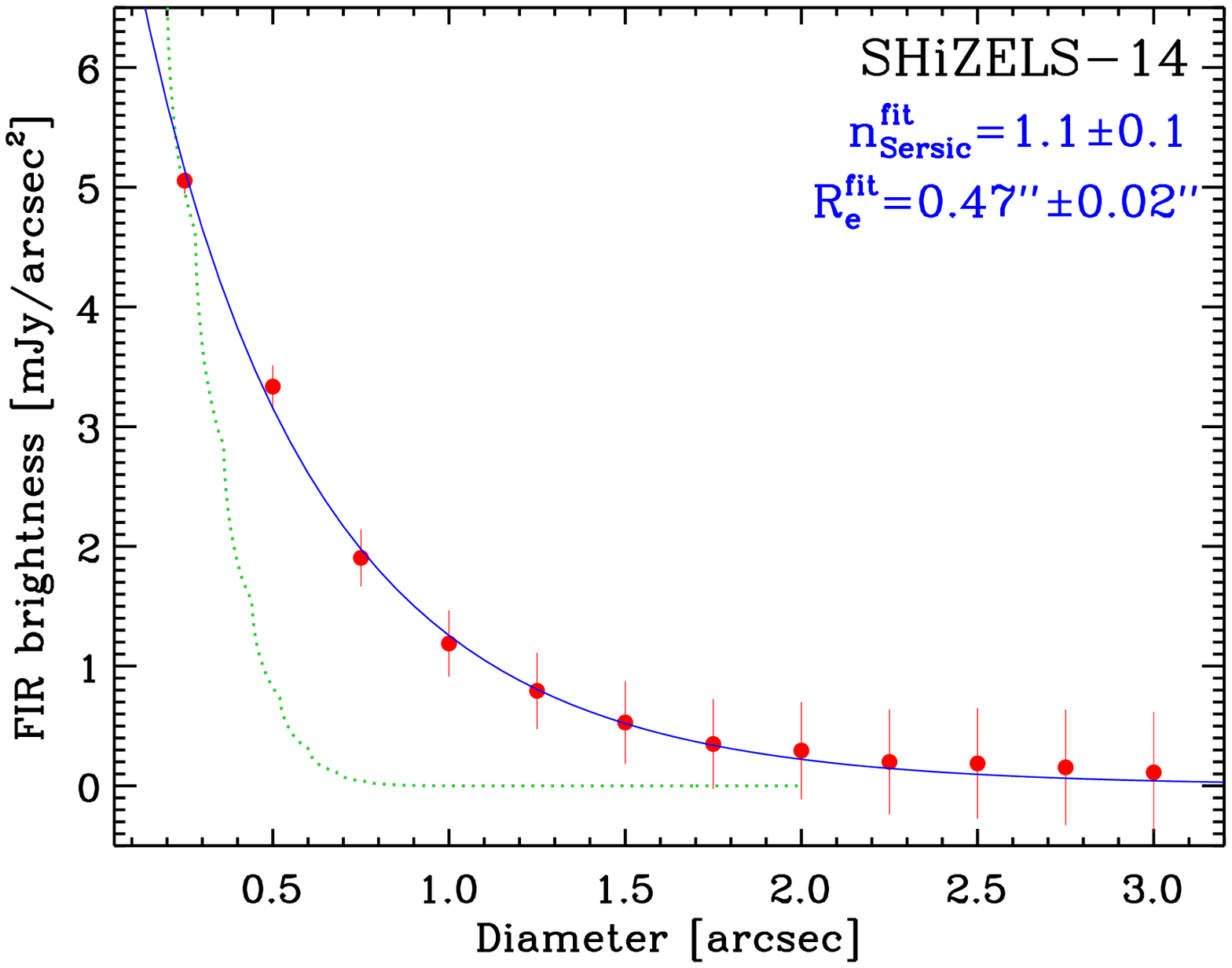}
\caption{The rest-frame $355\,\mu m$ surface brightness profile of the ALMA-detected
galaxies. The interval of each aperture annuli is $0\farcs25$ with the aperture cantered
at the peak pixel of the ALMA-tapered images. The green dotted
line shows the surface brightness profile of the ALMA synthesised beam. We can see
that continuum emission extends up to typically $\sim 2''$, i.e.\ about 16\,kpc in
diameter. The blue lines are the Sersic function fitting results. We denote the
fitting results in each panel.}\label{dust-profile}
\end{figure*}

We derive the dust continuum surface brightness profile by using the peak flux pixel
as the centres of the tapered ALMA images. Then we perform aperture photometry 
by using a series of ring apertures in steps of $0\farcs25$ (similar to the 
synthesised beam) in the natural-weighted ALMA images. Then we derive the surface 
brightness in each annulus for all of the four ALMA-detected galaxies, and 
show the results in Fig.~\ref{dust-profile}. We also show the surface brightness 
profile of the ALMA synthesised beam, normalised to the maximum value for each 
source in Fig.~\ref{dust-profile}. We fit the ALMA surface brightness profile 
with a Sersic function, and present the results in each panel. Sersic indices n 
of the ALMA images are lower than $n=2$, implying a disk-like morphology of the 
dust emission.
Based on the aperture photometry in each apertures, we also obtained the flux 
growth curve of each ALMA-detected galaxy, and derive the half-light radius
in Tab. \ref{table2}.

\section{ALMA non-detected targets}
\subsection{Individual properties}
\begin{itemize}
\item {\bf SHiZELS-21}, also known as UDS-10 in \citet{Molina2017}. This galaxy is a compact rotating galaxy. The H$\alpha$ velocity dispersion profile is double peaked at about $1.5$\,kpc from the galaxy center. The {\it HST} F606W image shows a double UV core structure which is barely resolved in the F140W image.

\item {\bf SHiZELS-8}: This galaxy is dominated by rotation and present clumpy features \citep{Swinbank2012a, Gillman2019}. None of the three H$\alpha$ clumps is detected by ALMA. Its metallicity is relatively low ($12+\log(\rm O/H)<8.3$) and the metallicity gradient is flat \citep{Swinbank2012a}. A simple dynamical description using carbon monoxide is presented by \citet{Molina2019}. The {\it HST} F606W image shows an extended UV morphology, while the F140W image shows a compact core in the galaxy. The clumpy structures seen in the H$\alpha$ map are not recovered by the F606W image.

\item {\bf SHiZELS-10}: This galaxy is compact in H$\alpha$ ($\sim$\,2.3\,kpc) and identified as a merger \citep{Swinbank2012a, Gillman2019}. The ALMA continuum emission is undetected, probably due to limitations in the surface brightness. On the other hand, the {\it HST} F140W, F606W and the VLT/SINFONI images show a good spatial consistency with bright dots and a long tail. 

\item {\bf SHiZELS-2}: The H$\alpha$ IFU observations evidence a clear rotation curve, including two clumps at the centre with a separation of $0\farcs2$ \citep[about 1.5\,kpc][]{Gillman2019}. The fainter clump locates in the rotation center. The orientation of the {\it HST} morphology is similar to that revealed by rotation in H$\alpha$. The {\it HST} images show a compact stellar and star formation distribution. The {\it HST} F606W image also shows that the star formation in the galaxy center is bright in rest-frame UV \citep{Gillman2019}.

\item{\bf SHiZELS-3}: The H$\alpha$ velocity map shows a clear rotation feature, while the velocity dispersion map shows two peaks separated by $\sim$2.5\,kpc, suggesting a complex dynamics probably associated to merging activity \citep{Gillman2019}. {\it HST} images show that the stellar mass and rest-frame UV star formation morphologies are compact.
\end{itemize}

\subsection{Stacking analyses}

\begin{figure*}
\centering
\includegraphics[width=0.95\textwidth]{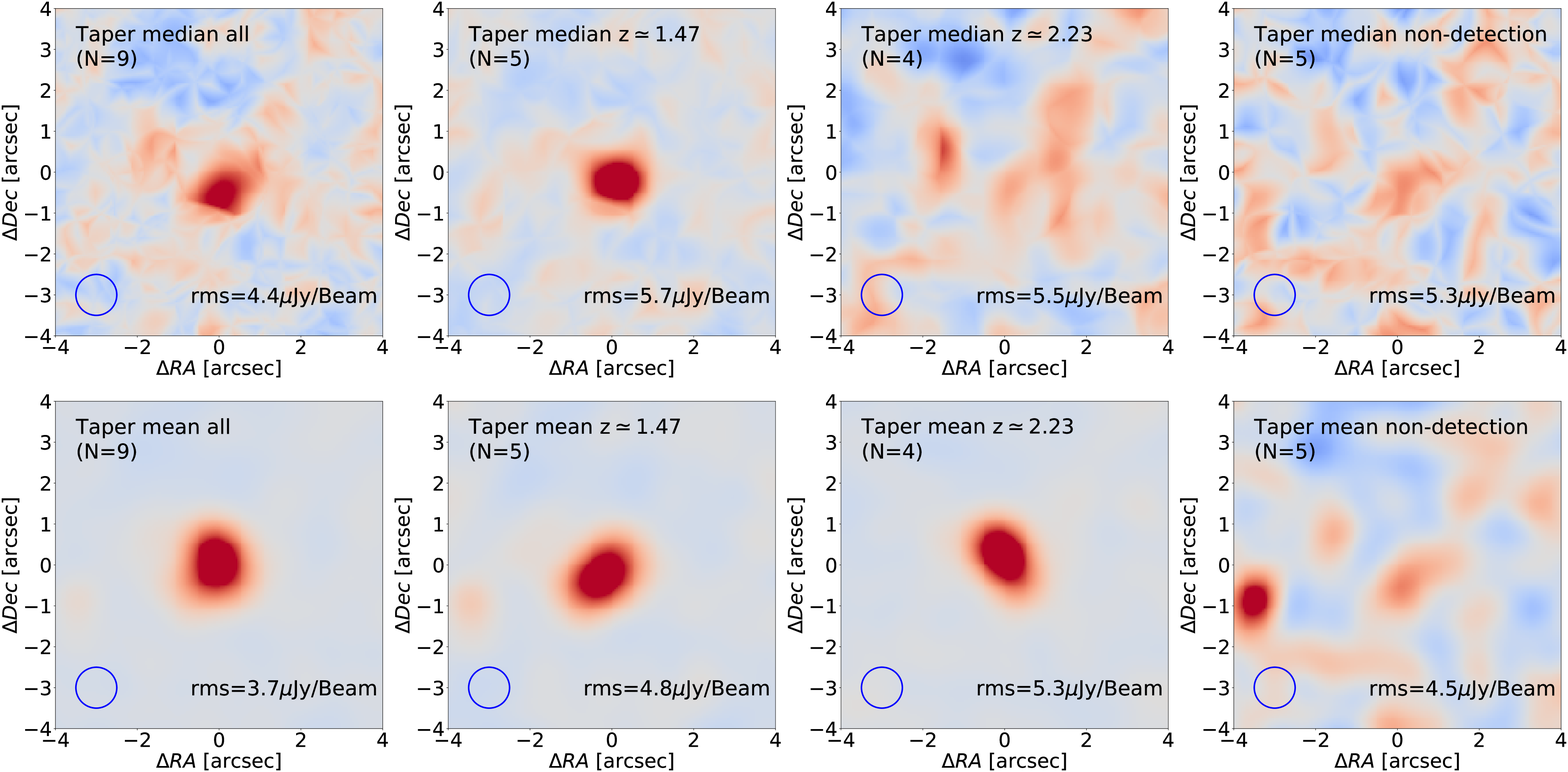}
\caption{
Postage stamps ($8''\times8''$) for the stacked continuum tapered (FWHM\,$=1''$) ALMA images. Upper and lower panels are 
the median and average stacks. From left to right are the stacks for different populations: all sources, only $z=1.47$, only $z=2.23$, 
all non-detected images (mixing galaxies at $z=1.47$ at $z=2.23$).}\label{stack}
\end{figure*}

In this appendix we explore the possibility to extract information from the non-detected
sources via a stacking approach. To do this, we stack the ALMA tapered image which are generated at
$1\farcs0$ resolution. We generate postage stamps (of $8''\times8''$) for each ALMA continuum image,
centred at the optical RA and Dec., to stack them based on median and average statistics (see Fig.~\ref{stack}).
We reach rms values from 3.5 to 5.5\,$\mu$Jy\,beam$^{-1}$ in these stacks.
At $z=1.47$, we detected three galaxies out of five, and both the average and median
show a significant emission at the image centre. For the $z=2.23$ population, only one out of four
targets is detected, hence the clear detection in the mean stack is clearly biased by the brightest galaxy.
This significant detection disappears when we look at the median estimate.
We also combined all non-detections together (mixing $z=1.47$ and 2.23 galaxies) in the right
panels of Fig.~\ref{stack}.

To estimate the significance of the stacks of non-detected ALMA images, we use a peak to noise criterion. 
The peak values are obtained from a 2-D Gaussian profile using a fixed centre and fixed width (FWHM) at 1$''$ 
(assuming point like detections). We measure the peak flux densities of 12.7\,$\mu$Jy\,beam$^{-1}$ for the mean stacked
image, and 13.1\,$\mu$Jy\,beam$^{-1}$ for the median one. Comparing these values with the background noise,
these peaks are only at $\sim$3$\sigma$ significance. We consider these stacks as non-detections.

Based on the rest-frame frequencies for these stack measurements, and considering the Rayleigh–Jeans
tail \citep[at 850$\mu$m flux; Equ. 16 in ][]{Scoville2016}, we can derive global ISM mass limits
for our SHiZELS targets. Considering median stacks, and 5$\sigma$ upper
limits for the $z=2.23$ population, we derive ISM masses of $\log(M_{\rm ISM}/M_\odot)=9.5$ at $z=1.47$
and $\log(M_{\rm ISM}/M_\odot)<9.2$ at $z=2.23$.

\section{Serendipitous ALMA detection}\label{APC}

\begin{figure*}
\centering
\includegraphics[width=0.3\textwidth]{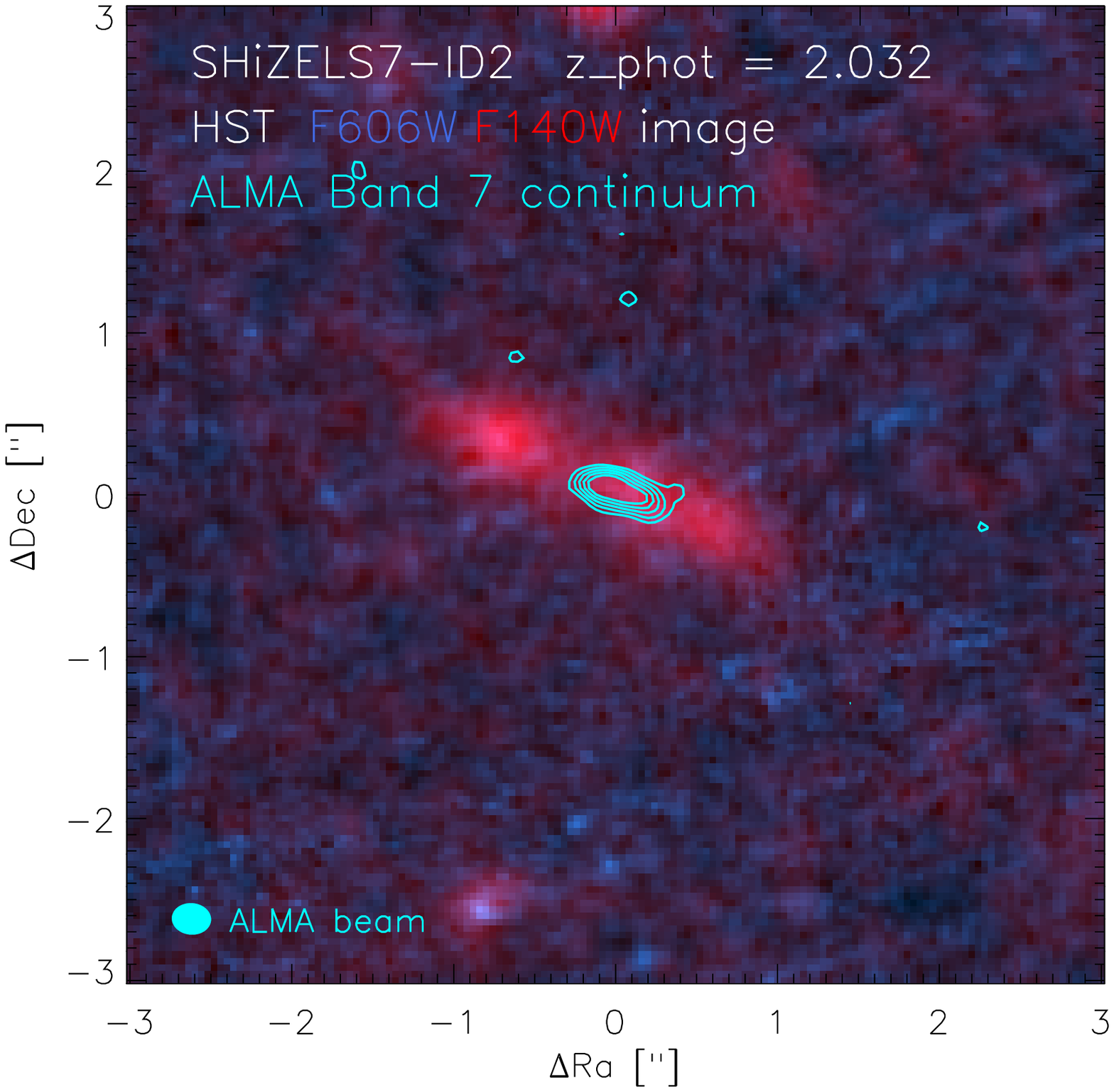}
\includegraphics[width=0.65\textwidth]{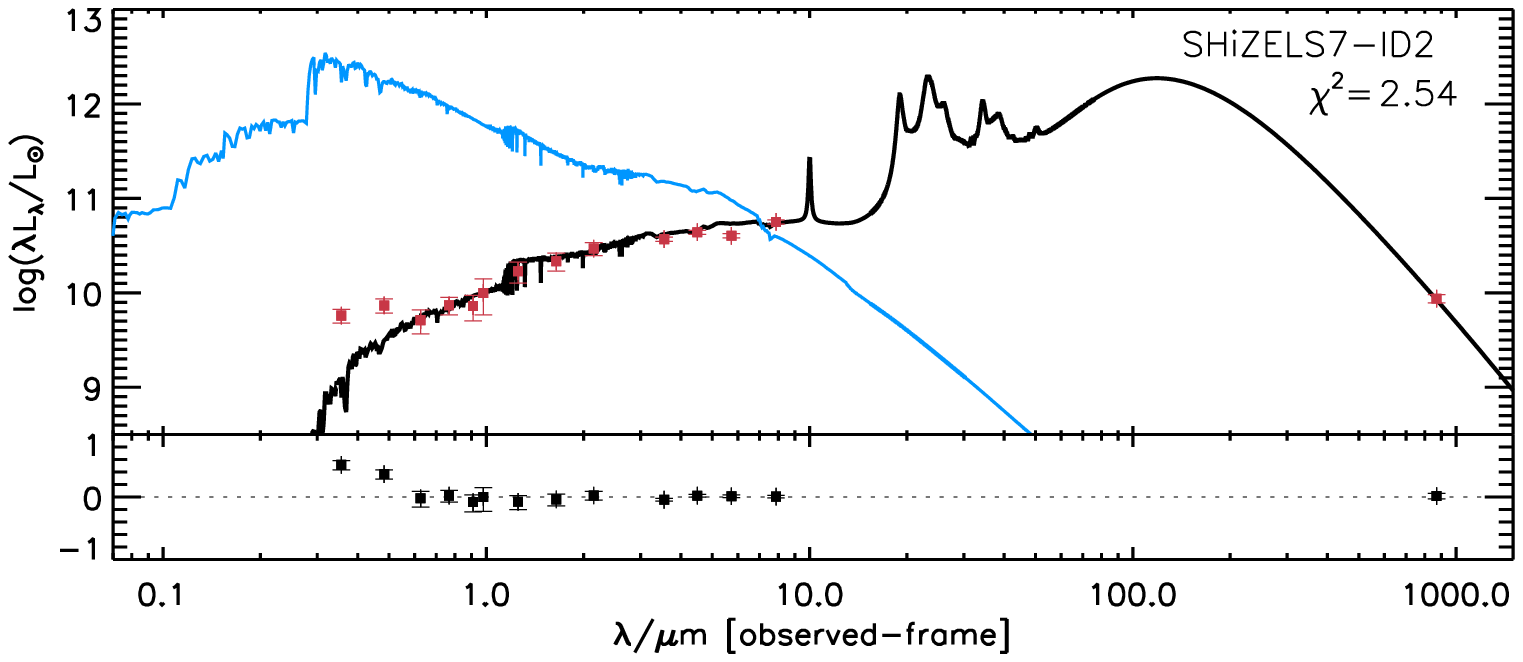}
\includegraphics[width=0.3\textwidth]{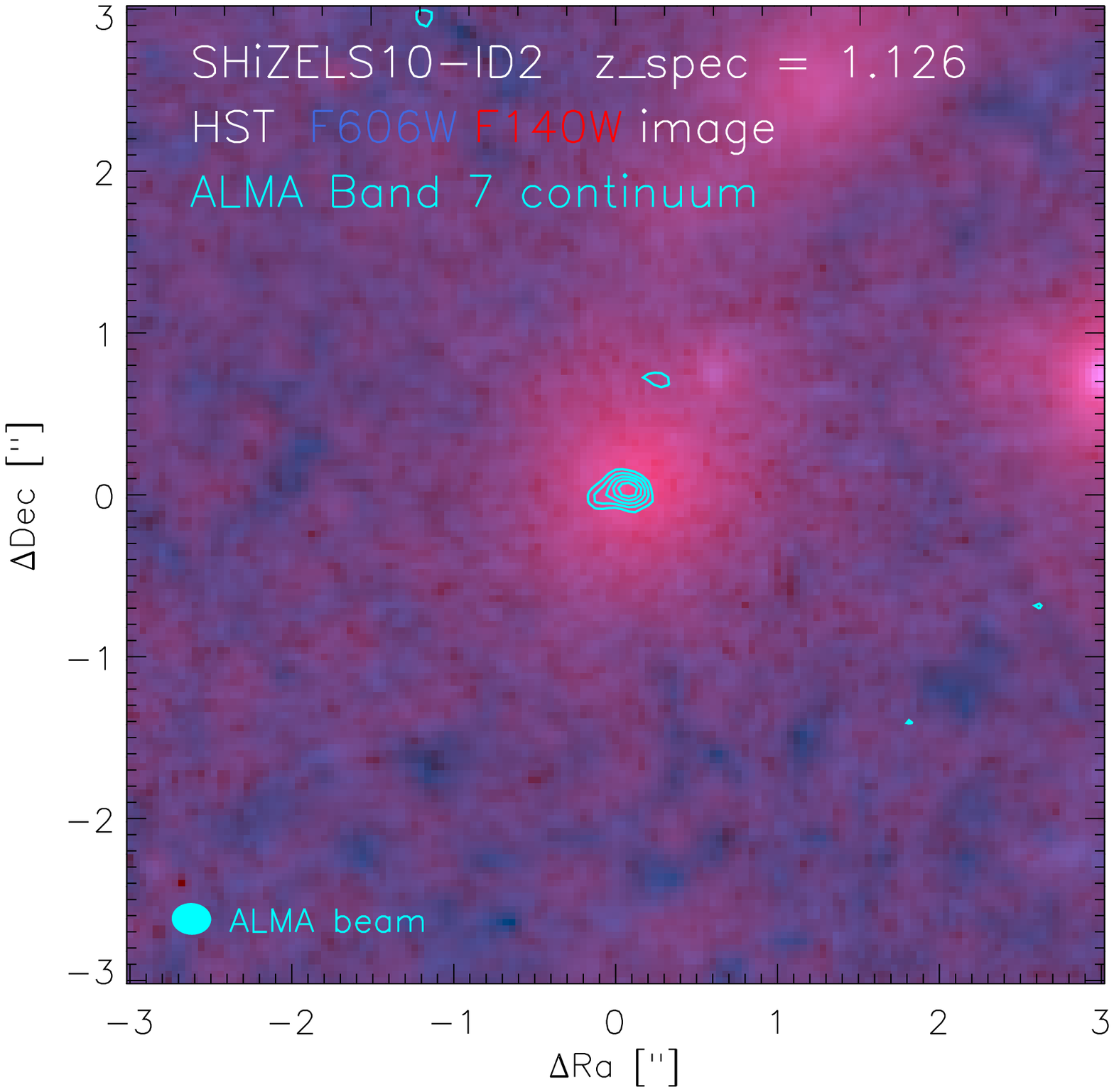}
\includegraphics[width=0.65\textwidth]{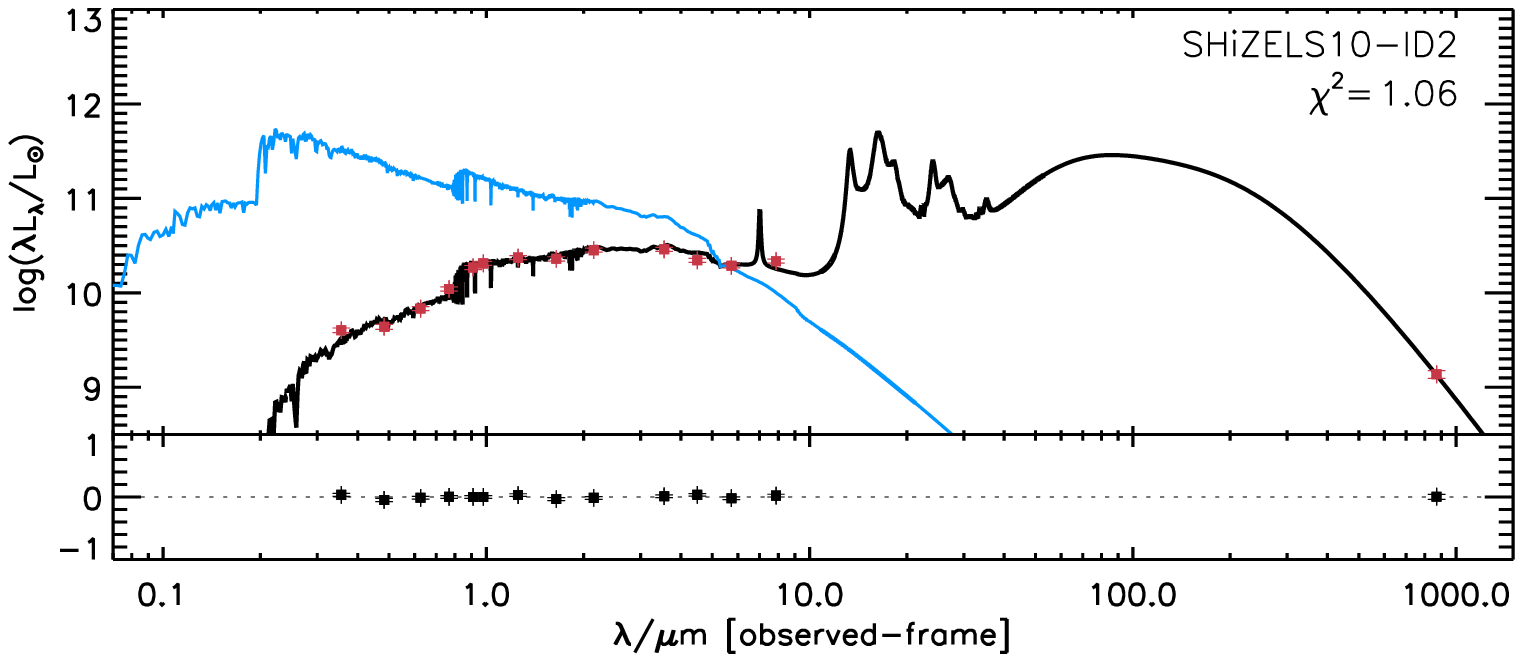}
\caption{Left panels: Fake color image (blue for F606W and red for F140W) with ALMA contour (3, 4, 5, 6, 7 $\times$rms). 
Right panels: Redshifts and SEDs taken from \citet{Mehta2018} as well as our new ALMA observations are shown in red dots. 
We fit the SED using MAGPHYS \citep{2008MNRAS.388.1595D}. The initial stellar spectra are shown in blue lines, and the 
model SED after considering the dust extinction are shown in black lines.
} \label{color}
\end{figure*}

In this appendix, we briefly characterise the properties of the serendipitous galaxies
SHiZELS7-ID2 and SHiZELS10-ID2. Both sources have been identified in the Multi-wavelength Photometric 
Catalog of the {\it Spitzer} Large Area Survey with Hyper-Suprime-Cam (SPLASH) in the Subaru XMM-Newton 
Deep Field (SXDF) \citep{Mehta2018}. SHiZELS10-ID2 presents a spectroscopic redshift at $z_{\rm spec}=1.126$ 
while SHiZELS7-ID2 only has a photometric redshift estimate at $z_{\rm phot} =2.03$. In Fig.~\ref{color} 
we show the ALMA continuum 355\,$\mu$m rest-frame contours on top of a fake colour optical image.
The ALMA flux densities are considered for a SED fitting approach including $U, G, R, I, Z, Y, J, H, Ks$ 
photometric bands taken from \citet{Mehta2018}. We derive the stellar mass by MAGPHYS
and the properties of these two targets are presented in Table~\ref{tabplus} and Fig. \ref{color}. 
Considering the ALMA flux, our stellar masses are consistent with the previous results within 0.5\,dex.
More properties of these two targets can be found in \citet{Mehta2018}.

\begin{table*}
\centering
\scriptsize
\caption{Properties of the serendipitous ALMA detections found in the field
of view of the SHiZELS galaxies presented in this work.}
\label{tabplus}
\begin{tabular}{lcccccccccc} 
\hline
ID     &    ID   &  RA (J2000) & Dec (J2000)   & Redshift & ALMA flux (mJy) & $\log(M_*/M_\odot)$ & $\log(M_*/M_\odot)$\\
       &  From \citet{Mehta2018}  &  &                &          & 873$\mu$m       & By MAGPHYS & From \citet{Mehta2018} \\
\hline            
SHiZELS10-ID2 & 954698 & 02:17:39.261 & $-$4:44:42.33 & $z_{\rm spec} = 1.126$          & 0.20 $\pm$ 0.03 & 10.4 $\pm$ 0.1 & 10.2 \\
SHiZELS7-ID2  & 874393 & 02:16:59.969 & $-$5:01:53.49 & $z_{\rm phot} = 2.033\pm 0.045$ & 0.32 $\pm$ 0.03 & 10.3 $\pm$ 0.1 & 10.6 \\
\hline
\end{tabular}
\end{table*}

\label{lastpage}
\end{document}